\documentclass[manuscript, acmsmall]{acmart}
\AtBeginDocument{%
  }

\usepackage{graphicx}
\usepackage{multirow}
\usepackage{algorithm}
\usepackage{amsmath}
\usepackage{url}
\usepackage{subcaption}
\usepackage{bbding}

\setcopyright{acmlicensed}
\acmJournal{TOPS}

\newcommand{\etal}{et al.}

\newcommand{\SRQ}[1]{\textbf{SRQ#1}}

\definecolor{color1}{RGB}{31,119,180}
\definecolor{color2}{RGB}{255,127,14}
\definecolor{color3}{RGB}{44,160,44}
\definecolor{color4}{RGB}{214,39,40}

\newcommand{\revised}[1]{#1}

\newcommand{\ATone}{AT1}

\newcommand{\AND}{AND}

\newcommand{\OR}    {OR}

\newcommand{\ATlabel}[2]{b_{{#2}}}

\def\showimages{1}
\def\showtables{1}

\def\showComments{1}

\if\showComments1
    \newcommand{\OG}[1]{{\color{blue} OG: #1}}
    \newcommand{\NS}[1]{{\color{purple} NS: #1}}
    \newcommand{\COM}[1]{{\color{red} #1}}
\else
    \newcommand{\OG}[1]{}
    \newcommand{\NS}[1]{}
    \newcommand{\COM}[1]{}
\fi

\begin{document}

\title[Attack Tree Distance]{{Attack Tree Distance}: A Practical Examination of Tree Similarity Measurement within Cybersecurity}

\author{Nathan Daniel Schiele}
\email{n.d.schiele@liacs.leidenuniv.nl}
\orcid{0000-0003-1186-1503}

\author{Olga Gadyatskaya}
\email{o.gadyatskaya@liacs.leidenuniv.nl}
\orcid{0000-0002-3760-9165}
\affiliation{%
  \institution{Leiden Institute of Advanced Computer Science}
  \city{Leiden}
  \country{The Netherlands}
}








\renewcommand{\shortauthors}{Schiele and Gadyatskaya}
\begin{abstract}
    Attack trees are a popular threat modeling method. In practice, there is often a need to compare attack tree models produced by human experts, based on both the structure of the tree and the meaning of the node labels. In this work, we investigate the problem of comparing attack trees and measuring their similarity. We define five different measures for measuring the distance between two attack trees: Label Distance (LD), Tree Edit Distance (TED), Radical Distance (RD), Multiset Distance (MSD) and Weighted Sum Distance (WSD). 
    We further propose a repeatable method of both theoretical and experimental attack tree distance measures validation. Our theoretical validation consists of a series of basic transformations to evaluate the behavior of distance measures with respect to specific types of transformations that may appear between two attack trees. To experimentally validate our distance measures, we designed and executed a human study ($n=39$) to collect a dataset of attack trees to be used for evaluation and comparison of the measures. 

    From our theoretical and experimental results, we find that applying semantic similarity as a means of comparing node labels is a valid approach. Further, we find four of the five attack tree distance measures are valid approaches in certain, varying circumstances. Our results suggest that these methods can already be used to identify similar real-world attack trees. Overall, this work lays the groundwork for improved threat model analysis, validation of AI-generated attack trees, and future research into threat similarity measurement in cybersecurity.
    
\end{abstract}

\ccsdesc[500]{Security and privacy~Formal security models}
\ccsdesc[500]{Security and privacy~Usability in security and privacy}
\ccsdesc[100]{Security and privacy~Security requirements}

\keywords{Attack Defense Trees, Threat Models, Tree Distance, Semantic Similarity}


\maketitle

\section{Introduction}
\label{sec:introduction}

\emph{Threat modeling} seeks to organize threat information that aids in comprehension and analysis of cyber attacks~\cite{andersonSecurityEngineeringGuide2020,schneierSecretsLiesDigital2000}. \emph{Attack trees} (ATs) are a popular method for threat modeling in software engineering~\cite{shostack2014threat,tarandach2020threat}, security requirements elicitation~\cite{rashid2016discovering,mai2018modeling}, and security risk assessment~\cite{ingoldsby2010attack,paul2014unifying}. ATs represent attack scenarios as labeled rooted trees, where the root label captures the high-level attacker's goal, and the labels of all other nodes represent particular sub-goals or steps to achieve them. There is a rich literature on attack tree models designed by experts to better understand threats to a particular system, e.g., ATMs~\cite{fraile2016using}, smart cars~\cite{kong2018security,ren2011novel}, cloud systems~\cite{wang2012threat,duncan2019combined}, smart grids~\cite{beckers2014determining,mclaughlin2010energy}, medical implants~\cite{siddiqi2018attack} or SCADA systems~\cite{ten2007vulnerability}. Furthermore, in the past few years, there has been a substantial interest in the automated synthesis of attack trees~\cite{konsta2024survey} using formal system models~\cite{widel2019beyond,ivanova2015attack,vigo2014automated,pinchinat2015atsyra,gadyatskayaRefinementAwareGenerationAttack2017,sowka2023review,konsta2024survey,audinot2018guided,paul2014towards}, libraries of attack patterns~\cite{jhawar2018semi,bryans2020template,bryans2023formal, haddadAutomatedMappingCVE2023}, or even machine learning methods~\cite{sowka2021towards, kekulComparisonAnalysisSoftware} and, most recently, large language models (LLMs)~\cite{gadyatskaya2023chatgpt,stridegpt2024,long2024}.

Given this proliferation of attack tree models produced for different scenarios, especially when they are synthesized automatically, novel research \revised{problems} emerge. \emph{How can we establish whether two attack trees produced by different experts or with different synthesis methods describe similar attack scenarios?} For example, consider the recent work by Nicoletti et al.~\cite{nicoletti2024hard} that proposes a method to represent MITRE attack campaigns as attack trees and compares these campaigns based on computed quantitative attributes; it would be very useful to quantify the similarity of different campaign models to identify threat actors that have related modi operandi. \revised{In his foundational paper, Schneier advises creating a library of reusable attack trees that could capture relevant attack scenarios, thus diminishing the need to have security experts around~\cite{schneierAttackTrees1999}. Researchers have started to crowd-source attack trees for benchmarking~\cite{dorfhuber2024quadtool}. In such a setting, when there is a collected library of attack trees, an automated way to identify already present attack scenarios similar to the newly submitted one will help categorize and curate new additions.}  
A related novel research \revised{problem} was \revised{investigated} in~\cite{gadyatskaya2023chatgpt}: \emph{how to establish the quality of an LLM-synthesized attack tree by comparing it to a tree designed by a human expert?} However, currently, there is no standard method for comparing attack trees or measuring their similarity. 
Existing attack tree semantics can partially answer this question as semantics provide a way to define the meaning of an attack tree model~\cite{mauwFoundationsAttackTrees2006}. However, a semantics-based approach can only conclude whether two \revised{attack} trees are equivalent: the established attack tree semantics do not allow estimating similarity between these trees or comparing tree models with similar yet unequal labels that might appear identical to a human~\cite{mauwFoundationsAttackTrees2006}. Consider, for example, two attack goals \emph{rob a bank} and \emph{steal money from a bank}. To a human expert, these two attack goals might appear quite similar. A strict semantics-based approach, e.g., \emph{\`a la} Mauw and Oostdijk~\cite{mauwFoundationsAttackTrees2006}, will consider these to be two different attacks. We thus aim to propose new methods that would \revised{automate} the assessment of AT similarity.  





\textbf{Research question and contributions.} 
Our fundamental research question (RQ) is: \emph{How do we measure the distance between two human-created attack trees?} 
Examining this question leads to three sub-research questions:
\begin{enumerate}
    \item \SRQ{1}: How to compare human-created sets of attack tree node labels?
    \item \SRQ{2}: How to compare the structure of attack trees?
    \item \SRQ{3}: How to validate a distance measure for attack trees?
\end{enumerate}

To answer these questions, 
we first propose a method for comparing the labels of attack trees based on semantic similarity using SBERT~\cite{devlin2018bert,reimers-2019-sentence-bert}. This is then refined into our first suggested distance measure called \emph{label distance}, which estimates how close the whole sets of labels of two trees are. \textbf{Our first contribution is a repeatable method for comparing two lists of labels based on semantic similarity}.

We then apply the proposed \emph{label distance} and develop an approach for comparing the structure of ATs. We propose three additional methods for calculating the distance between two attack trees that incorporate structural elements of the tree: (1) \emph{tree edit distance}~\cite{zhang_editing_1992};  (2) \emph{radical distance} inspired by~\cite{schiele2021novel} and the refinement semantics discussed in~\cite{gadyatskayaRefinementAwareGenerationAttack2017}; and (3) \emph{multiset distance} that stems from the multiset attack tree semantics proposed in~\cite{mauwFoundationsAttackTrees2006}. We further propose the fifth method, which is a \emph{weighted combination of these three distance measures}. \textbf{Our second contribution is these four methods for comparing the structure of two attack trees}.

Following this, we provide a method for both theoretically and experimentally validating any attack tree distance measurement method to ensure that the distance measure reflects how best understanding of how two trees differ. Specifically, we establish a complete set of basic transformations that a tree could undergo, and assess our distance measures with respect to these transformations. We also collect a series of human-drawn ATs and compare them using our distance measures. Finally, we experiment with multi-dimensional scaling~\cite{torgerson1952multidimensional} as another approach to demonstrate the validity of the proposed methods. \textbf{Our third contribution is a repeatable method for validating any attack tree distance measure}.

We believe that in the future, when we can expect attack models to be synthesized using dedicated generative AI methods~\cite{elshareffacilitating,happe2023getting,xu2024autoattacker,zhang2024attackg,zhang2024vtt}, similarity measurement for ATs will become an even more important problem. Our work paves the way to solve this problem and to develop other, even more robust attack tree similarity measurement methods. 

\textbf{The structure of this paper.}
We first introduce some background concepts about attack trees (Section~\ref{sec:background}) and discuss the related literature on tree distance metrics (Section~\ref{sec:related-work}). 
We then define an idea of \emph{semantic similarity} for node labels that allows matching realistic, ``\emph{noisy}'' labels and relaxes the strict label equality requirement (Section~\ref{sec:semantic-similarity}). We conceptualize and define 4 distance measures for attack trees (Section~\ref{sec:distance}).
Subsequently, we propose our validation methodology (Section~\ref{sec:validation}) and we report the results of evaluating these measures following our validation method, (Section~\ref{sec:results}). We discuss our findings in Section~\ref{sec:discussion} and conclude with Section~\ref{sec:conclusion}.

\section{Background}
\label{sec:background}



\if\showimages1
\begin{figure*}
\resizebox{0.92\linewidth}{!}{
    \includegraphics{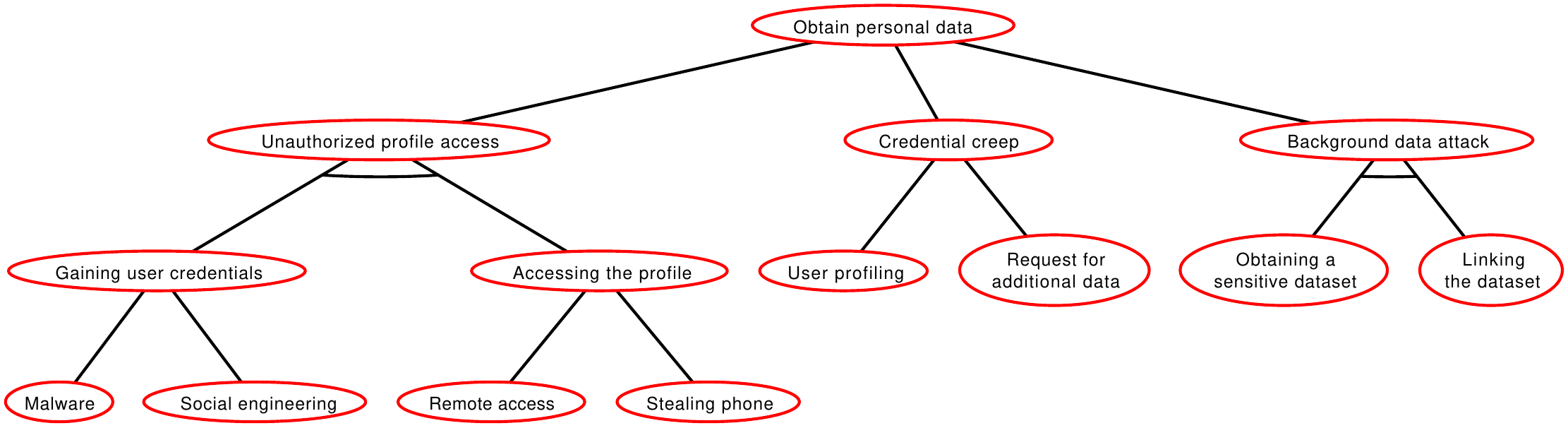}
}
\caption{An attack tree adapted from Naik~\etal~\cite{naikEvaluationPotentialAttack2022} that is used in the study described in Section~\ref{ssec:method-study-design}. }
    \label{fig:tartgetAT}
    \Description{Attack tree showing a categorical breakdown of an attack}
\end{figure*}
\else
\fi


We define attack trees (ATs), as originally described by Bruce Schneier~\cite{schneierAttackTrees1999}, to be a rooted acyclic structure with the following recursive definition from~\cite{gadyatskayaRefinementAwareGenerationAttack2017}.

%
An attack tree $t$ is defined as:
\[
    t = b | b\Delta(t_1, ..., t_i)
\]

Where $b \in \mathbb{B}$ is an action or goal in service of an attack, considering $\mathbb{B}$ to be the set of all possible attack components. For our purposes, $b$ is considered to be the label of a node. $\Delta$ is the refinement operator of a node, defining the relationship between child nodes. In this work, $\Delta$ is either \AND\ or \OR.

Traditionally, the root node of an attack tree represents the main attacker's goal, and the lower nodes represent \emph{refinements} of the main goal into various attack steps and options available~\cite{mauwFoundationsAttackTrees2006}. The \AND\ refinement means that all children nodes must be achieved in order to succeed at the parent node level; while the \OR\ refinement means that the attacker has options, and achieving any of the children nodes is enough to succeed at the parent node level. The lowest nodes in the tree that are not refined are called leaf nodes. 

\revised{An example of an attack tree describing how personal data can be obtained is available in Figure~\ref{fig:tartgetAT}. The root node is \emph{Obtain personal data}, which is refined using an \OR\ operator into three sub-goals: \emph{Unauthorized profile access}, \emph{Cred. creep}, and \emph{Background data attack}. The first sub-goal of \emph{Unauthorized profile access} is further refined using an \AND\ operator into two options: \emph{Gaining user credentials} and \emph{Access profile}. The first of these is further refined using an \OR\ operator into two steps: \emph{Malware} and \emph{Social engineering}. The other nodes are not further refined and are thus leaf nodes. Each node is labeled with a short textual description of the attack step or goal it represents. The specificity of the labels can vary, and will fundamentally depend on the application context and the user designing the tree.}

We note that the above 
definition is \emph{syntactic}: following the original formulation by Schneier~\cite{schneierAttackTrees1999}, it only defines a well-formed attack tree. Subsequently, researchers defined several semantics for attack trees to formally model these structures. For example, the seminal work by Mauw and Oostdijk defined two semantics: the propositional semantics, capturing an attack tree as a Boolean formula, and the multiset semantics that defines an attack tree to be a set of multisets representing combinations of leaf nodes that independently enable a successful attack~\cite{mauwFoundationsAttackTrees2006}. 

Semantics attempt to represent the meaning of an attack tree and enables equality testing: if two trees can be represented by the same Boolean formula or the same set of multisets, then they are equal in the considered semantics. If two trees are semantically equivalent, then all quantitative analyses with attributes compatible with the semantics will result in equal values~\cite{mauwFoundationsAttackTrees2006}. Trivially, this semantics-based reasoning can be extended from equality to testing for inclusion (being a subset). 
However, it is not evident how to extend the semantics-based approach to measuring similarity or distance between two attack trees. 

Indeed, to consider the similarity of attack trees, we need to take into account two key aspects of this model: \emph{node labels} (including labels of intermediate nodes that are absent from, for example, the multiset or propositional semantics~\cite{mauwFoundationsAttackTrees2006}) and the \emph{refinement structure} that defines what attacks are subcomponents of which other steps. Moreover, as we mentioned, in real life, node labels are somewhat \emph{noisy}: experts (or LLMs~\cite{gadyatskaya2023chatgpt}) can propose slightly different textual labels to define the same underlying attack idea. Thus, to be able to compare real attack trees (which we also call \emph{``unfiltered''} in this work), we need to develop means to account for this noise.

\section{Related Work}
\label{sec:related-work}

\subsection{Natural Language Processing (NLP) and Semantic Inference}
\label{ssec:related-work-nlp}

Natural language processing is the subfield of computer science focused on machine processing of human-generated language~\cite{chowdhary2020natural}. There are multiple subfields within NLP focusing on different avenues of machine analysis of human language~\cite{cambriaJumpingNLPCurves2014, hirschbergAdvancesNaturalLanguage2015}. For example, syntactic analysis focuses on leveraging the structure of language to understand meaning~\cite{sidorovSyntacticNgramsMachine2014}. Semantic analysis, on the other hand, focuses on the meaning of language~\cite{kapetaniosNaturalLanguageProcessing2013}.

There are multiple methods to process human language semantically. A common method involves converting individual words to semantic meaning vectors using a pre-trained Recurrent Neural Network (RNN) using Word2Vec developed by~\cite{mikolov2013efficient}.  Palmer~\etal\ extended syntactic analysis by adding semantic role labels (information concerning predicates)~\cite{palmerPropositionBankAnnotated2005}; a work later expanded upon by Collobert~\etal\ who created a neural network using, in part, semantic role labels to comprehend whole sentences~\cite{collobertNaturalLanguageProcessing}. Han~\etal\ proposed a method for extracting semantic features of words and classifying them using a Long-Short-Term Memory (LSTM) network~\cite{hanNovelMalwareDetection2021}, while Turian~\etal\ devised a semi-supervised approach to achieve the same goal of semantic representation of words~\cite{turianWordRepresentationsSimple}. Kalchbrenner~\etal\ created a Convolutional Neural Network (CNN) to model the semantic meaning of sentences~\cite{kalchbrennerConvolutionalNeuralNetwork2014}, which uses tree-structured sentence representations to make semantic comparisons. However, there is no clearly established method to definitively compare the meaning of sentences~\cite{chowdhary2020natural}. \revised{Chandrasekaran and Mago peformed a literature survey of semantic similarity techniques in literature, exploring the different methods used to calculate the semantic similarity between two text elements \cite{chandrasekaranEvolutionSemanticSimilarity2022}. Our research most closely aligns with the Explicit Semantic Analysis (ESA) type of Corpus-based semantic analysis. This type of semantic similarity creates a vector based representation of text (in our case node labels) and then compares those vectors with cosine similarity.}

Natural Language Sentence Mapping (NLSM) is a specific research area within NLP that focuses on comparing sentences with each other and predicting matching sentences~\cite{luMKPMMultiKeywordpair2022, wangBilateralMultiPerspectiveMatching2017}. NLSM matches the challenge we tackle in comparing the labels of two trees, discussed in detail in Section~\ref{sec:semantic-similarity}. Hu~\etal\ uses a CNN to compare sentences~\cite{huConvolutionalNeuralNetwork2014}, while  Wang~\etal\ proposes a multi-perspective matching model to do the same~\cite{wangBilateralMultiPerspectiveMatching2017}. Several works have expanded BERT specifically for the task of sentence matching. Zhou~\etal\ extends BERT to introduce relation awareness~\cite{zhouRelationAwareEntityMatching2022}, while Liu~\etal\ introduces syntax and semantic awareness to BERT~\cite{liuSentenceMatchingSyntax2020}. Although many of these approaches would be easily applicable for our goals to establish semantic similarity of AT labels (Section~\ref{sec:semantic-similarity}), given the lack of a clearly established applicable method for NLSM~\cite{liuStructuredAlignmentNetworks2018}, we seek an approach that is both easily reproduced and iterated upon (for example, by using either a pre-trained model or a completely different, task-specific model).

One of the most critical developments in semantic word representation is the development of Bidirectional Encoder Representations from Transformers (BERT) from  Devlin~\etal~\cite{devlin2018bert}. BERT is a transformer-based model that is pre-trained on a large corpus of text. It is capable of understanding the context of a word in a sentence and can be fine-tuned for a variety of NLP tasks. Liu~\etal\ replicated and improved the results of BERT in RoBERTa~\cite{liuRoBERTaRobustlyOptimized2019}. 

We use the Sentence-Bidirectional Encoder Representations from Transformers (SBERT) from Reimers and Gurevych~\cite{reimers-2019-sentence-bert}. This is a state-of-the-art sentence embedding model built upon the well-established BERT model. We use several pre-trained SBERT models for our semantic analysis. 
We take inspiration from Tracz~\etal\ who used BERT in a product matching solution~\cite{traczBERTbasedSimilarityLearning2020}. Their contribution was the novel product matching solution, not the application of BERT. Similarly, our contribution is the distance measures, not the application of SBERT in the distance measures.

\revised{\subsection{Semantic Similarity of Taxonomies (Information Content)}}
\label{ssec:related-work-taxonomies}

\revised{
Philip Resnik used information content (IC) within taxonomies to compare the semantics (underlying representation of taxonomy) of the taxonomies themselves~\cite{resnikUsingInformationContent1995}, which is similar to our goal of comparing ATs. 
Derived from this work, Jiang and Conrather ~\cite{jiangSemanticSimilarityBased1997}, and Dekang Lin \cite{linInformationTheoreticDefinitionSimilarity1998} both explore IC as a mechanism to compare words within taxonomies. Giuseppe Pirro provides the groundwork for incorporating multiple features into an IC similarity~\cite{pirroSemanticSimilarityMetric2009}, similar to how we incorporate multiple features into our distance measures.
}

\revised{
Jiang and Conrather develop an experimental comparison methodology which helps inform our experimental design~\cite{jiangSemanticSimilarityBased1997}. 
De Nicola~\etal\ provides a parametric similarity method for taxonomies, as well as a comparison of different similarity methods, which further helps inform our experimental design~\cite{denicolaParametricSimilarityMethod2023}.
}

\revised{
Meng~\etal\ introduced a hybrid model using semantic (meaning) inference from WordNet~\cite{mengNewHybridSemantic2012}
In a similar vein, Jia Xiao also introduced a hybrid model which used IC to improve the semantic similarity measure between different words~\cite{xiaoHybridModelImprove2024}. 
These works help inform ours, as we incorporate the semantic similarity of sentences (AT labels) into our distance measures, and we selected BERT as our mechanism to compare sentences.
Finally, De Nicola~\etal\ provides a detailed overview of a comparison of different semantic similarity methods~\cite{denicolaInformationContentMethods2025}, these methods can inform how we can compare different measures that we develop, and provide insight into both how our measure can potentially be improved and how our experimental design can be further developed}.

\subsection{Tree Edit Distance}
\label{ssec:related-work-ted}

Distance between data structures is not a new concept. Many works have explored the idea of ``distance'' between strings. In string edit distance, the difference in strings is given by a min-cost path taken by either adding a character, removing a character, or replacing a character. By finding the minimum cost needed to transform one string into another, a ``distance'' value can be given. The cost of transformation increases with the difference between two strings~\cite{yujianNormalizedLevenshteinDistance2007, masekFasterAlgorithmComputing1980}.

Tree edit distance (TED) is seen as an extension of the string edit distance. Tai first tackled the problem, suggesting an edit distance metric for two directed acyclic graphs (DAGs) \cite{tai_tree--tree_1979}. Zhang and Shasha have written the seminal work on tree edit distance~\cite{Zhang_Shasha_1989}. We describe this distance and our application of it in more detail in Section~\ref{ssec:ted}.


Most of the research and development on tree edit distance focuses on optimizing computations. As shown by Zhang~\etal, the tree edit distance problem for unordered trees is an \textit{NP}-Complete problem~\cite{zhang_editing_1992}. As such, the development of novel optimal calculation strategies is necessary to enable the comparison of larger tree structures. Yoshino \etal\ developed a dynamic programming A$^*$ algorithm for computing unordered tree edit distance (UTED), which offers significant performance gains over exhaustive search~\cite{yoshino_dynamic_2013}. Their method uses an A$^*$ algorithm to construct a search tree of mappings. The distance is then calculated from these mappings. Unfortunately, these mappings rely on an absolute equivalence between nodes of two trees to establish a mapping, which cannot be established a priori for ``unfiltered'' (real-world) attack trees, and thus cannot be applied in our case.  Other works based on the A$^*$ algorithm methodology, such as the optimizations offered by Paassen~\cite{paasen_-algorithm_2021} for computing UTED will have the same drawback. However, the intuition to first find a potential optimal mapping and then find the distance based on the mapping is a useful methodology that we can apply to our work.

McVicar~\etal\ have developed a method of calculating UTED in SuMoTED by focusing on allowing nodes to move up or down a tree and organizing edits around a consensus tree~\cite{mcvicar_sumoted_2016}. This methodology allows them to achieve polynomial-time calculation of UTED. However, the application of this methodology to attack trees presents difficulties. Namely, we have the same issue as the A$^*$ algorithm: the requirement of exact equivalence between node labels.


Pawlick and Augusten proposed RTED, a more optimal method of calculating ordered TED (OTED)~\cite{pawlik_rted_2011}. Their methodology is faster than Zhang and Shasha~\cite{Zhang_Shasha_1989}, but specifically for larger trees (500+ nodes). Attack trees, by contrast, do not grow this large as they become cognitively too demanding as threat models~\cite{andersonSecurityEngineeringGuide2020}. Thus, as such, RTED does not offer a significant advantage for attack trees over Zhang and Shasha~\cite{Zhang_Shasha_1989}.





\subsection{Tree Kernels}
\label{ref:tree-kernels}
A common way to calculate tree distances in machine learning is the use of tree kernels. Tree kernels compare tree structures by computing some positive-definite function over a subset of features of the trees (for example, the dot product). There is no commonly accepted method for selecting features or selecting the positive-definite function~\cite{shinComprehensiveStudyTree2014}. A distance measure must already be clearly defined in order to build a tree kernel~\cite{shinTheorySubtreeMatching2015}. Moschitti proposed a mechanism for applying tree kernels to NLP by representing sentences with syntactic parse trees and comparing those trees~\cite{moschittiMakingTreeKernels2006}; Collins and Duffy proposed a similar method for comparing parse trees~\cite{collinsConvolutionKernelsNatural2002}. Tree kernels have been used for semantic analysis, such as by Aseervatham~\etal~\cite{aseervathamSemanticKernelSemistructured2007}. 

Building on Collins and Duffy, Kashima and Koyamada proposed a method of comparing trees by comparing the substructures of the trees~\cite{kashimaKernelsSemiStructuredData}. This method is based on the idea that the similarity of two trees is the sum of the similarities of their substructures. Shin built upon the work of Kashima and Koyamada by comparing trees by matching substructures, a matching methodology similar to what we employ~\cite{shinTheorySubtreeMatching2015}. These methods informed our development of the radical distance, described in Section~\ref{ssec:rd}. In almost all cases, tree kernels either only compare structural information or are used to represent non-tree structured information (such as sentence structure). 
Tree kernels have been used for both NLP and structural tree comparison, but have not been used for both simultaneously; our work is essentially this. To our knowledge, no previous work has attempted to apply tree kernels to unfiltered tree-structured data. It may be possible to develop a tree kernel that can both compare the structure of the tree and the meaning of the node labels (\textit{e.g.} a nested tree kernel), but this is outside the scope of this work.

\subsection{Applications of Tree Distances}
\label{ssec:related-work-treedist-application}

Tree distance applications appear in many fields. As already mentioned, tree edit distance is an important component of NLP and is used in many different NLP works to achieve machine comprehension, with Palmer~\etal\ comparing tree representations of semantic roles~\cite{palmerPropositionBankAnnotated2005}, Kalchbrenner~\etal\ comparing syntax trees~\cite{kalchbrennerConvolutionalNeuralNetwork2014}, and Liu~\etal\ comparing tree representation of sentences~\cite{liuStructuredAlignmentNetworks2018}.

In other fields, for example, tree (edit) distance has been used by Otu and Sayood to compare phylogenetic trees in biology~\cite{otuNewSequenceDistance2003}, by Chen and Zhang to compare RNA secondary structures~\cite{chenImprovedAlgorithmTree2014}, by Xu~\etal\ to detect fake web sites by comparing tree representations of the DOM structure~\cite{xuDetectingFakeSites2016}, by McNamee to compare patent classifications (stored in hierarchical tree structures)~\cite{mcnameeCantSeeForest2013}. A work from Lakkarju~\etal\ especially informs our work, as it shows a mechanism to compare XML files based on tree distance (creating tree representations of XML tags, but not the content of the tags)~\cite{lakkarajuDocumentSimilarityBased2008}. We believe that our work could fill a similar gap, or offer a pathway toward filling this gap, in the field of threat modeling.

\subsection{Applications of Distance Measures in Cybersecurity}


Distance measures are an often discussed topic within cybersecurity; however, distances of threat modeling is not a well explored area.
The most directly related work to ours is by Lo~\etal\, which used a variety of distance measures to determine insider threats within a CERT dataset~\cite{loDistanceMeasurementMethods2018}. One of their distance measures included the Levenshtein distance that we explore in Section~\ref{ssec:comparing-labels}. Related to this is a work by Rashid~\etal, who used Hidden Markov Models to accomplish the same  \cite{rashidNewTakeDetecting}. A common discussion of distance comes from discussions of security metrics, as described by Andrew Jaquith: if security is quantified into metrics, these metrics can be compared with distance measures~\cite{jaquith2007security}. This was more closely examined by Krautsevich~\etal, proposing several mathematical formalisms for determining security hierarchies~\cite{krautsevichFormalApproachSecurity}. Similar work to Krautsevich has been performed by Wagner~\etal, who focus on using distance measures for privacy comparisons~\cite{wagnerTechnicalPrivacyMetrics}. Measuring the difference between Common Vulnerabilities and Exposures (CVEs) and Common Weakness Enumeration (CWEs), and mapping CVEs and CWEs to malicious behavior is another active area of research in cybersecurity~\cite{haddadAutomatedMappingCVE2023, kekulComparisonAnalysisSoftware, heAutomaticallyIdentifyingCVE2024}. 
To our knowledge, we are the first to examine distance as it is applied to threat modeling and attack trees.

\revised{
    \subsection{Other Threat Models}
    \label{ssec:related-work-other-threat-models}
    While attack trees are a prominent threat modeling technique~\cite{shostack2014threat}, other threat modeling techniques are also widely used and studied. Most closely related to attack trees are fault trees, from which attack trees were designed~\cite{schneierAttackTrees1999}. Fault trees represent faults using logical structures, similar to how attack trees represent attacks~\cite{lallieEmpiricalEvaluationEffectiveness2017}. Another graphical threat modeling technique is attack graphs, which is a state based representation of attacks~\cite{sheynerAutomatedGenerationAnalysis2002a}, where nodes represent system states and edges represent actions that transition the system from one state to another. 
    Finally, CORAS is another graphical threat modeling technique which represents attack scenarios by classifying different elements of the attack and visually showing the relationships between these elements~\cite{solhaugCORASLanguageWhy2014}. 
}


\textbf{Novelty of this work.} To the best of our knowledge, there are no proposals in the literature on how to compare attack trees or to measure their similarity, or to compare tree-structured data with noisy, human-made labels. This is the gap we are aiming to close with this work. 

\section{Semantic Similarity}
\label{sec:semantic-similarity}

In this section, we address our first research question, \textbf{SRQ1}, on how to compare and measure distance between noisy attack tree labels and sets of such labels.

\subsection{Label Sets}


As there are no methods for comparing real-life tree-structured data, we develop our own approach. Following \cite{bengioLabelEmbeddingTrees2010}, which focuses on specific node labels to help analysis, and \cite{heDecompositionTreeSpatiotemporal2015}, which suggests a mechanism to decompose tree structures, we propose a robust and repeatable method for comparing unfiltered data that is organized in a tree structure. We propose to compare the sets of labels of the tree structure to form a mapping similar to the suggestion of Zhang and Shasha~\cite{Zhang_Shasha_1989}. This mapping can be used in lieu of absolute equivalence. We create the label sets by extracting all node labels from an AT. As such, a comparison of the two label sets will amount to a comparison of the node labels without any structural information. Thus, if we were to have two trees with identical structures but different node labels, the comparison of these two label sets should be a sufficient proxy for the comparison of the trees those sets were generated from.

\subsection{Comparing Labels}
\label{ssec:comparing-labels}

Fundamental to comparing label sets is comparing the labels within those sets. We discussed much of the state-of-the-art of language processing in Section~\ref{ssec:related-work-nlp}. We start from a more basic premise: directly comparing the labels as strings. A common method of comparing strings is to use string edit distance (Levenshtein distance)~\cite{levenshtein1966binary,rizaNaturalLanguageProcessing2020}. Levenshtein distance will count the number of changes needed to be made to the strings to convert one string to another. This causes larger values for longer strings (as more edits are needed). However, this introduces a problem: longer strings will tend to have larger distances than shorter strings due to the increased number of comparisons.

The Levenshtein distance can be normalized by dividing the resulting distance value by the largest string length (a process previously discussed in \cite{marzal1993computation, yujianNormalizedLevenshteinDistance2007}), yielding a percentage measurement of the number of characters in a string that match. This allows for a direct comparison of two labels as strings which yields a value between 0 and 1 regardless of the length of the strings. 
The following equation yields the normalized string edit distance, for two labels $\ATlabel{d}{i}$ and $\ATlabel{e}{j}$:

\[
\delta\left(\ATlabel{d}{i}, \ATlabel{e}{j}\right) = 1 - \frac{\text{Levenshtein}\left(\ATlabel{d}{i}, \ATlabel{e}{j}\right)}{\max\left(\left|\ATlabel{d}{i}\right|, \left|\ATlabel{e}{j}\right|\right)}
\]

\if\showimages1
\begin{figure}
    \centering
    \resizebox{0.95\linewidth}{!}{%
    \includegraphics{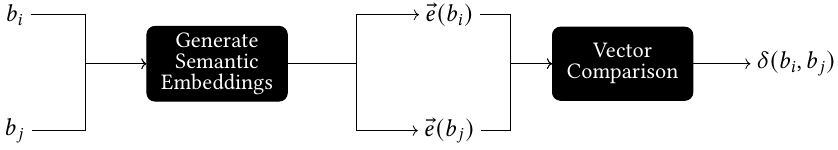}
    }
    \caption{Process of calculating the distance between two node labels.}
    \label{fig:semanticreplacement}
    \Description{Steps of how the difference between two labels is calculated. This description is present in the text.}
\end{figure}
\else
\fi

However, Levenshtein distance does not consider the meaning of the labels, which is more fundamental to how we would like to compare the labels. We wish to develop measures which can reflect how similar two attack trees are in \emph{meaning} rather than simple string distance. As such, we opt for comparing the semantic meaning of the labels using SBERT~\cite{reimers-2019-sentence-bert} to create a vector representation of the labels. We can then compare these vectors using cosine similarity, which is a measure of the angle between two vectors~\cite{li2013distance}. This gives us a value between 0 and 1, and is directly comparable to using the normalized Levenshtein similarity. The process of calculating the semantic similarity is shown in Figure~\ref{fig:semanticreplacement}. We further explore the negative effects of using Levenshtein distance in Section~\ref{ssec:method-alternatives}.

\subsection{Comparing Label Sets}

With an established mechanism to compare two labels, we move on to the problem of comparing sets of labels. We do this by taking the Jaccard distance between two sets of labels~\cite{kosub2019note}. The Jaccard distance is a measure of similarity between two sets, and is calculated as the size of the intersection of the two sets divided by the size of the union of the two sets~\cite{gonzalez-huitronJaccardDistanceSimilarity2023}. However, the Jaccard distance requires a binary determination of an equivalence between two labels. We thus first use the semantic similarity measure we defined above to determine the similarity of two labels, and then compare this similarity against a threshold $\epsilon$. We discuss this threshold in the next section.

\subsection{Equivalence Threshold}
\label{sssec:threshold-problem}

Equivalence is a binary measurement: labels are either equivalent or not. To ensure comparability of distance measures, all of the AT distance measures we conceptualize further in Section~\ref{sec:distance} require binary determinations of equivalence between pairs of labels. While it may be possible to establish distance measures that allow for more nuanced determinations of similarity (\textit{i.e.} not binary), this is outside the scope of this work. We are interested in establishing a mechanism to compare distances between attack trees and validate those measures. We leave the refinement of attack tree distance measures to future work.

However, the semantic similarity we defined above is a continuous value between 0 and 1, which by its nature is not binary. To convert this continuous measure of semantic similarity to one of semantic equivalence, we employ a threshold value of $\epsilon$. Two labels with any semantic similarity value above $\epsilon$ are considered equivalent, and labels with a semantic similarity value below $\epsilon$ are considered not equivalent.

Thus, we encounter the threshold problem: the problem of defining a threshold $\epsilon$ for the semantic similarity between two labels. This threshold is necessary to determine if two labels are similar enough to be considered the same. We further propose a way to determine a suitable range for the threshold based on empirical results with our attack tree datasets. Currently, the same threshold $\epsilon$ is applied to all pairs of labels. In the future, it might be interesting to explore whether more nuanced interpretations of semantic similarity (e.g., varying the threshold for different label types) enhance the attack tree similarity measurement approaches. 

In Section~\ref{sssec:finding-epsilon}, we discuss the process of finding the right $\epsilon$ for a given dataset. This process involves iteratively examining the dataset to establish what equivalences need to be assumed in order to create a meaning distance measure. From these equivalence assumptions, an optimal $\epsilon$ can be established. This process is similar to the process of establishing a threshold for a binary classification problem, where the threshold is set based on the desired false positive and false negative rates~\cite{santosoGeneticProgrammingApproach2018}. Our exact process is visually represented in Figure~\ref{fig:finding-epsilon}.

\subsection{Label Distance}
\label{ssec:label-distance}

We now define the Label Distance (LD) -- a distance measure between the sets of labels of two attack trees; LD is functionally a Jaccard distance on two sets of labels.  We derive this measure from the mappings suggested in Tai~\cite{tai_tree--tree_1979} and Zhang and Shasha~\cite{Zhang_Shasha_1989}, with an algorithm influenced by the A$^*$ tree edit distance algorithm from Yoshino~\etal~\cite{yoshino_dynamic_2013}. A pre-trained SBERT model is used to calculate the distance between all label pairs, following the semantic similarity comparison introduced in Section~\ref{sec:semantic-similarity} as shown in Figure~\ref{fig:semanticreplacement}. 

We then create an $M \times N$  matrix $D$, which represents all the distances between all label pairs for two trees. The largest value at index $i, j$ of this matrix is taken, and the corresponding row ($i$) and column ($j$) are removed from $D$, while recording the mapping between $b_i$ and $b_j$. This is repeated until all nodes are included in a mapping set. If the largest value in $D$ is below some value $\epsilon$, we consider the labels to be different, and a direct mapping is not created. We add the nodes as individuals to the mapping (mapping to/from $\Lambda$, indicating that these labels would be removed or added). We then calculate the cost of each node that would be removed or added, considering these operations at the unit cost of 1. \revised{This process is shown in Figure~\ref{fig:ld_flowchart}. The total cost of the mapping is the distance between the two sets of labels.}

The normalized measure is calculated by taking this distance value and dividing it by the number of nodes in the largest tree. \revised{As Jaccard distance follows the triangle inequality~\cite{kosub2019note}, it follows that LD follows the triangle inequality. As LD is additive, it cannot be negative; further, the distance between identical sets is 0. Therefore, LD is a metric.}

\if\showimages1
\begin{figure}
    \centering
    \resizebox{0.6\linewidth}{!}{%
    \includegraphics{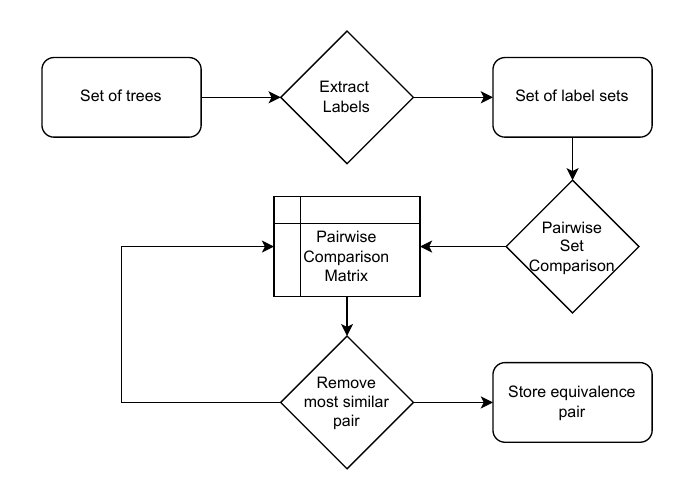}
    }
    \caption{\revised{Flowchart of the steps in label distance}}
    \label{fig:ld_flowchart}
    \Description{Flowchart showing how label distance works}
\end{figure}
\else
\fi

This distance measure considers all the node labels within an attack tree. It uses a semantic comparison to examine the meanings of the nodes, which would allow for this distance measure to work on unfiltered trees. However, this measure does not consider the structure of the tree in any way and, therefore, does not represent how the nodes are organized. If we examine the attack tree in Figure~\ref{fig:tartgetAT}, for an attack tree with all the same label names, but all nodes directly attached to the root (a single root node with 13 child nodes with identical names as in Figure~\ref{fig:tartgetAT}), the resulting label distance would be zero. Thus, as such, label distance offers a way to quickly examine if the meanings of the nodes are similar, but does not offer a way to examine the structure of the tree.

Applying LD to the examples provided in Figure~\ref{fig:dist-example}, we would first create two lists of labels of both Figures~\ref{fig:dist-example-1}~and~\ref{fig:dist-example-2}. Then we could compare these lists according to their semantic similarity along the process described in Section~\ref{sec:semantic-similarity}. The labels that are identical will be matched first, then the labels will be matched according to their semantic similarity in decreasing order. This is repeated until only the node ``brute force'' from Figure~\ref{fig:dist-example-2} remains. Provided all other nodes match (given a sufficiently permissive value of $\epsilon$), LD will result in a distance of 1 (normalized distance of 0.167).

\section{Distance Measures}
\label{sec:distance}

We have established a mechanism for comparing node labels and using this mechanism to compare the label sets of two attack trees. However, these comparisons do not take into account the structure of the attack trees; we address including structural information in this section. In each of the following distance measures, we make use of the same semantic similarity measurement we use for the label distance to compare node labels. These results will answer our research question \textbf{SRQ2} about providing means of comparing the structure of attack trees, while also considering the label similarity. As a running example to demonstrate how these proposed distance measures work, we refer to the two similar attack trees shown in Figure~\ref{fig:dist-example}. 



\if\showimages1
\begin{figure*}
    \centering
    \begin{subfigure}[b]{.38\linewidth}
        \resizebox{\linewidth}{!}{
           \includegraphics{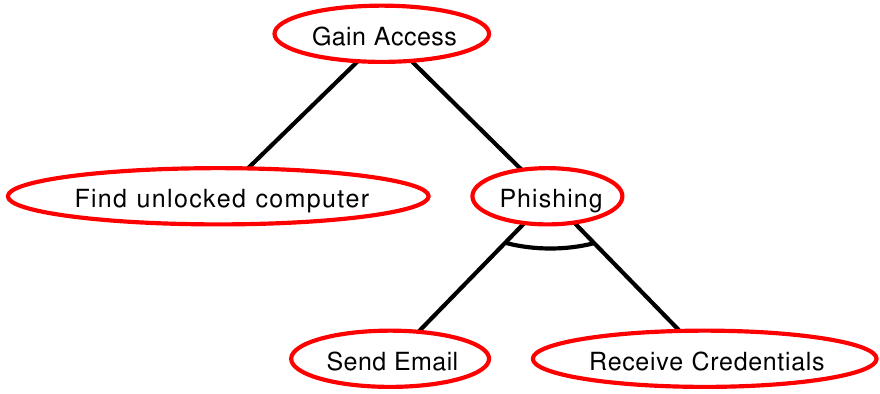}
        }
        \caption{A simple attack tree.}
        \label{fig:dist-example-1}
    \end{subfigure}
    \begin{subfigure}[b]{.48\linewidth}
        \resizebox{\linewidth}{!}{
           \includegraphics{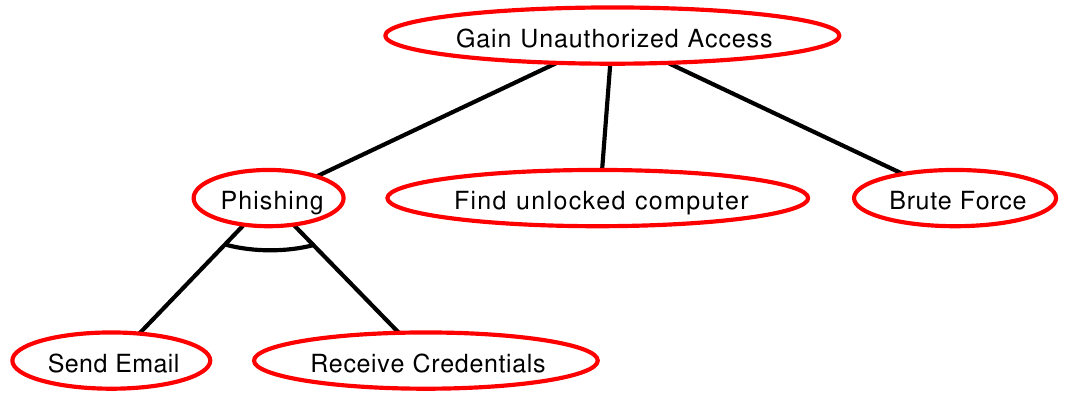}
        }
        \caption{A slightly modified attack tree.}
        \label{fig:dist-example-2}
    \end{subfigure}
    \caption{Two similar attack trees with small differences to model the distance measures.}
    \label{fig:dist-example}
    \Description{Two attack trees}
\end{figure*}
\else
\fi




\subsection{Tree Edit Distance}
\label{ssec:ted}

Tree edit distance (TED) is a measure of the difference between two trees by defining an optimal edit path; that is, what changes are needed to turn one tree into another. The seminal work in this field is by Zhang and Shasha~\cite{Zhang_Shasha_1989}. This was developed from, and is highly based on, the previously discussed string edit distance (Levenshtein distance)~\cite{tai_tree--tree_1979}. In their work, they describe a simple algorithm for calculating the distance between two trees~\cite{Zhang_Shasha_1989}. This algorithm is based on the idea of a \textit{forest} distance, which is the distance between two forests, or sets of trees. The result of this algorithm is a measure of the costs of edits needed to transform one tree into another, with the convention being that any edit requires a cost of 1. The possible edits are: 
\begin{itemize}
    \item matching, in which two nodes are defined to be equivalent, which is the only edit without cost, 
    \item insertion, in which a node is added, 
    \item deletion, in which a node is removed, and changing one node into another, in which a node has its label replaced. 
\end{itemize} 
For the edits, we need to know if two nodes are equivalent. This is done by comparing the semantic similarity of the two nodes; this is performed in the same manner as we did for the label distance. If the semantic similarity is above a given threshold, we consider the nodes to be equivalent; otherwise, we consider them to be different. As the TED value will provide us with a number of edits needed to transform one tree into another, we can normalize this value by dividing it by the number of nodes in the tree with the most nodes. This will give us a normalized value between 0 and 1, which is useful for comparing trees of different sizes. \revised{Li and Chenguang proved that the non-normalized TED is a metric~\cite{li2011metric}.}


The tree edit distance from Zhang and Shasha works on ordered trees, which for our purposes presents a challenge, as our attack trees are assumed to be unordered. Zhang~\etal\ showed that unordered tree edit distance is a MAX SNP-Hard problem~\cite{zhang_max_1994}. Our assumption that attack trees do not grow to be very large allows us to remain unconcerned about the processing time of unordered tree distance. 

We note that previous work on unordered tree edit distance has offered various polynomial-time approximate solutions, all making assumptions to speed up computations. Fundamentally, all make the same assumption: that there is absolute equivalence between nodes within two trees. Either nodes have identical labels or they do not. This would preclude their use on unfiltered data, as nodes are not guaranteed to have the same label, and indeed, the trees we may want to compare can have widely diverse label sets. On unfiltered data, these algorithms may find a local maximum instead of the global maximum needed to create a mapping (as is done for the label distance described in Section~\ref{ssec:label-distance}). As such, these algorithms are unsuitable for our purposes if applied directly.


Applying TED to the examples provided in Figure~\ref{fig:dist-example}, we follow the process described in \cite{Zhang_Shasha_1989}, by calculating key roots (the rightmost parent of each subtree) and then computing the forest distance. For the specific mechanics of this distance, we refer the reader to the work by Zhang and Shasha~\cite{Zhang_Shasha_1989}. The resulting distance, given a value of 0.7 or less for the semantic similarity threshold $\epsilon$, is 5 (normalized to 0.833). This result demonstrates that TED is unable to handle nodes that are reordered; TED treats these nodes as completed modified.

\subsection{Radical Distance (RD)}
\label{ssec:rd}

In our previous work on attack trees, we proposed a mechanism of attack tree decomposition that focused on the concept of \emph{radicals}, or subtrees consisting of a single parent, refinement, and set of children~\cite{schiele2021novel}. Indeed, attack trees are often developed sequentially, by iteratively refining one leaf node at a time~\cite{audinot2018guided}. We thus take this decomposition approach and define the radical distance as a way to compare the resulting set of radicals. This approach of comparing substructures has been done before, and relies on the assumption that mapping substructures can represent the overall structure~\cite{shinTheorySubtreeMatching2015,kashimaKernelsSemiStructuredData}.

In Radical Distance (RD), we first decompose an attack tree into a collection of subtrees that are each a unit height, indexed by each subtree's root node. We then perform a semantic comparison of each of the subtree roots; this semantic mapping is the same greedy mapping we conduct in label distance. We calculate the distance subtree by subtree, adding a distance of 1 if the root nodes are not equal, and a distance of 0.5 if the refinements are not equal. For the children, we perform a greedy matching operation similar to label distance. We do not include the distances of children that are themselves roots of other radicals, as these children will be evaluated when roots are compared in another iteration. This process repeats until all children are compared, added, or removed.

\revised{As RD is additive, it will always be positive. RD only results in a distance of 0 if the radical roots, refinement, and all child nodes are equal. As the node comparisons use the same mechanism as LD, the node comparisons fulfill the triangle inequality for RD. The refinement comparisons further fulfill the triangle inequality, as either all refinements are the same resulting in an added distance of 0, or a refinement must be changed twice, which would result in a distance of 1. As such, RD fulfills the triangle inequality and is therefore a metric.}


Applying RD to the examples provided in Figure~\ref{fig:dist-example}, we first compute the radical decomposition of both attack trees. These are the one height subtrees rooted at ``Gain Access'' and ``Phishing'' for Figure~\ref{fig:dist-example-1} and ``Gain Unauthorized Access'' and ``Phishing'' for Figure~\ref{fig:dist-example-2}. We then compute the semantic distance between the root nodes of each of the radicals, in a manner similar to LD, taking the most closely related radicals first. In this case, the two radical roots that are identical (``Phishing'') are selected first, and the radical distance is calculated by comparing the root nodes, refinements, and all children (excluding children that themselves are roots of other radicals). In this case, this comparison yields a difference of 0, given a value of 0.7 or less for the semantic similarity threshold $\epsilon$. This is repeated for the other two radicals, which have a distance of 1 (``brute force''), resulting in an overall distance of 1 (normalized to 0.167).

\subsection{Multiset Distance}
\label{ssec:msd}

Attack trees can be represented mathematically in many ways; one well-established attack tree semantics is the multiset semantics proposed by Mauw and Oostdijk~\cite{mauwFoundationsAttackTrees2006}. In the multiset semantics, each multiset represents a single complete attack vector within the attack tree, and a set of these multisets represents the whole attack tree. To define the multiset distance, we can use the Jaccard distance between two multisets. 

To expand the Jaccard distance to count some similar elements of multisets as identical, we need to calculate the semantic similarity between the elements of the multisets. This would result in a similar problem to the label distance problem or the modification of Zhang and Shasha tree edit distance, as we do not know a priori which labels should be matched with which ones. Because we assume that attack trees are reasonably small, we perform this calculation by calculating the Jaccard distance between each possible pair of sets, using semantic similarity to determine the equivalence of the elements in each set. This is functionally performing label distance on each pair of multisets. We then consider the two sets that have the smallest distance to be identical (a greedy mapping). We then remove these sets from the list of sets and repeat this process until all sets are compared.  
Ultimately, this results in a calculation that is functionally a nested Jaccard distance. 
\revised{For reasons similar to LD, means that MSD is a metric~\cite{kosub2019note}.}

Multiset semantics, like many mathematical attack tree semantics \cite{jhawar_attack_2015,mauwFoundationsAttackTrees2006}, contain only the leaf node labels. All other labels are reduced from the tree. Additionally, the exact structure of the tree is not generally reconstructible. Ultimately, semantics are meant to answer whether two trees are equivalent, which would mean that the exact structure of the tree is not necessarily important. In attack tree distance, however, the exact structure may be significant. By comparing the differences between multisets, we lose some information about the structure of the tree. We also lose the information contained in any intermediate nodes within the tree.

Applying MSD to the examples provided in Figure~\ref{fig:dist-example}, we first calculate the multiset semantic representations of both trees. These would be 
$\texttt{\{\{Send Email},$ 
$\texttt{Recieve Credentials,\}},$ 
$\texttt{\{Find}$ 
$\texttt{unlocked}$ 
$\texttt{computer\}\}}$ 
for Figure~\ref{fig:dist-example-1} and 
$\texttt{\{\{Send}$ 
$\texttt{Phishing}$ 
$\texttt{Email},$ 
$\texttt{Get}$ 
$\texttt{Credentials\}},$ 
$\texttt{\{Find}$ 
$\texttt{open}$ 
$\texttt{computer\}},$ 
$\texttt{\{Brute force\}\}}$ 
for Figure~\ref{fig:dist-example-2}. We then calculate the Jaccard distance between each possible pair of sets, using semantic similarity to determine the equivalence of the elements in each set. Starting from the pair of sets with the lowest Jaccard distance (the sets $\texttt{\{Find unlocked computer\}}$ and $\texttt{\{Find open}$ $\texttt{computer\}}$), we add the Jaccard distance of these sets (in this case 0) to the overall distance measure. These sets are then removed from the list of sets. This is repeated until all sets are compared, and any sets remaining (in this case $\texttt{\{Brute force\}}$) are compared to an empty set. In this case, this yields a distance of 1 (normalized to be 0.167), given a value of 0.7 or less for the semantic similarity threshold $\epsilon$.

\subsection{Weighted Sum Distance (WSD)}
\label{ssec:wsd}

The four distance measures that we considered previously (LD, TED, RD, and MSD) stem from some intuitions about how attack trees could be compared based on their components. We also consider it interesting to explore whether a combination of these measures would yield interesting results. We thus introduce the weighted sum distance (WSD) as a weighted average of the four baseline distance approaches discussed above. 

In the next section, we will first assess the four distance measures described in this section according to our experimental design described in Section~\ref{sec:validation}. We will then search for a combination distance measure that is the weighted sum of the other four distance measures. This weighted sum can be expressed as the following:
\[
    \alpha_1*\text{LD}+ \alpha_2*\text{TED} + \alpha_3*\text{RD} + \alpha_4*\text{MSD}
\]

For a given vector $\alpha = [\alpha_1, \alpha_2, \alpha_3, \alpha_4]$. We will look for an optimal alpha such that the performance of this weighted sum outperforms any individual distance measure shown above. \revised{As WSD is a sum of metrics, it is itself a metric.}

Applying WSD to the examples provided in Figure~\ref{fig:dist-example}, we would first calculate LD, TED, RD, and MSD for the two trees. We would then calculate the weighted sum of these values according to the selected value of $\alpha$ to find the WSD. The value of WSD will depend both on the value of $\epsilon$ (as it will affect the four primary distance measures) as well as the selected values of $\alpha$, which affect the sum.



\section{Validation Method}
\label{sec:validation}

In order to evaluate any approach for comparing attack trees, we must first establish a means of validating the results and assessing the quality of this approach. We propose two components for validating the distance measures.

\subsection{Experimental Validation}\label{ssec:exp-validation}
We offer an experimental design that can be used to validate a distance measure on ``real-world'', human-produced data. This requires a validation dataset that contains two sets of attack tree models. First, we require a base set of trees which are identical. This can be created by asking participants to create an attack tree describing the same precisely defined scenario. Any variation between the trees would likewise be representative of ``real-world'' label variation, and not due to differences in attack steps. In this way, we can evaluate how the distance measures handle similar but not fully identical data. 

Second, we require the base set of trees to be slightly expanded by the authors. This results in a second set of trees that are expected to be slightly different, but not very diverse. At the same time, the base set contains attack trees that are subtrees of this second set. This allows us to evaluate the behavior of the distance measures when comparing a tree to a known subset of that tree, and to compare a set of attack trees that are slightly different from each other. This behavior should be predictable, and we can assess distance measures if they display the behavior we expect. We describe the details of our study design and the data collection process below.


\subsubsection{Experimental Study Design}
\label{ssec:method-study-design}
We conducted an experiment with students who were assigned a project related to attack trees as a part of their regular coursework. The project was a mandatory, graded assignment. Students had the option to allow for their anonymized responses to be collected for research purposes (opt-in for the study). Our ethical considerations are described in more detail in Section~\ref{ssec:ethics}. 

The assignment consisted of several components that required students to create an attack tree using a web application we provided. 
The component of the study that provided data for validating our approach contained two parts. In Part A, students were tasked with creating an attack tree from a provided, written scenario; students were instructed to only and fully adapt the written scenario. 


The scenario text was created by reading the attack tree in Naik~\etal~\cite{naikEvaluationPotentialAttack2022} top-down (abstraction level by abstraction level), always going from left to right. Our intention with Part A was for participants to create trees that all reflect the same scenario in roughly the same manner. However, given that participants could elect to phrase nodes differently or include varying amounts of information within nodes, we expected that trees would be similar without being identical. Using this set of trees, we can evaluate how well the distance measures handle ``unfiltered'' node labels. We further refer to models created in this part as \textbf{AT1}.


We then asked students in Part B to expand upon their attack tree by ``doing your own research, add at least 5 new nodes and 2 new refinements to the attack tree you created in the previous section''. In all, this creates a relatively predictable yet varied dataset. In Part A, all trees should roughly be the same, bearing minor changes in the node labels or the order of nodes. We expect any large differences in Part A to be due to a misunderstanding of the assignment or misreading the scenario. As Part B is built from each students' Part A, we expect that the ``core'' of their attack tree to remain unchanged; the added nodes and refinement should introduce slightly larger trees that would still be somewhat similar to the original ones. We will refer to the extended attack trees produced in this part as \textbf{AT2}.

Finally, we asked students to create a completely new attack tree based on a scenario of their own choosing. These trees should be completely unique and are used in Section~\ref{ssec:results-mds} to validate how the distance measures we defined map similar trees close to each other. The set of completely unique trees is referred to as \textbf{AT3}.

Based on previous experience with similar studies, we expect most students to add exactly 5 new nodes and 2 new refinements, which should allow for some predictable distance variations (e.g., the cost of adding 5 new nodes in the tree edit distance). This allows us to evaluate the distance measures as a metric. As each student produced three trees, a base (AT1) and an extension of the base (AT2), and a completely unique tree (AT3), by comparing them we can validate the distance measures. If the distance measures are not able to recognize the base tree, the distance measure may not be suitable.

Further, we use multi-dimensional scaling (MDS) to map the distance of the trees to a two-dimensional space. This allows us to visualize the distance between the trees and perform manual checks to see if the trees that are mapped close to each other are actually similar.



\subsubsection{Participants}
39 students participated in this study conducted in the fall semester of 2023. They were all third-year bachelor students taking part in a minor on cybersecurity at Leiden University (NL). Most students were in policy-related majors such as Security Studies or International Relations. The participants, on average, had only a few months of programming experience. Students were asked if they had any prior knowledge of threat models or attack trees, and only a few students had any prior knowledge. 

\subsubsection{Ethics}
\label{ssec:ethics}

This study was approved by the Science Faculty Ethics Review Board at Leiden University. Students were informed of the study objectives and design, and gave written consent for their responses to be included in the study. The assignment provided to students was a mandatory, graded, course component, but participation in the study was \emph{opt-in} and fully voluntary. \revised{The ethical considerations of this study were very similar to those of a previous study we performed in~\cite{schiele2025limited}. For more details on the ethical considerations of this study, we refer to ethics considerations sections of Schiele~\etal~\cite{schiele2025limited}.}

\subsection{Theoretical Validation}
\label{ssec:methodology-examples}

This validation method involves defining a series of \emph{basic transformation examples} (BTEs) that represent the most fundamental ways in which two attack trees could differ from each other. The idea is that these BTEs only express the transformations that we would expect to see, or that we explicitly would desire to compare two distance measures along. This design has been used previously, with select examples used to model specific behavior of edit distances~\cite{tai_tree--tree_1979,akutsu_tree_2021}. In these works, select BTEs are used to model specific behavior of the proposed distance measures and to validate the distance measures.

We define a comprehensive set of BTEs to show all possible basic transformations. For each of these BTEs, we state our intuitive understanding of the distance, and then provide the distance according to our five distance measures. For example, if an order of \AND\ or \OR-refined children is changed, this does not affect the underlying attack scenario representation (as both of these refinement operators do not mandate any specific order for executing the attack actions). Thus, the distance between two such order-perturbed attack trees is intuitively expected to be 0. We then provide this BTE to the distance measures and compare it to our intuitive understanding.


We describe and show all of our BTEs in Figure~\ref{fig:counterexamples}. In all of our BTEs except for Figure~\ref{sfig:bte-order-reversed}, we expect the distance between Figure~\ref{sfig:base} and the provided BTE to be one. In Figure~\ref{sfig:bte-order-reversed}, we expect the distance to be zero.



\if\showimages1
\newcommand{\CEWidth}{.32\textwidth}
\newcommand{\ThreeVertSpace}{\vspace{0cm}}
\newcommand{\resizeSize}{2.5cm}
\begin{figure}
    \centering

\begin{subfigure}[b]{\linewidth}
        \centering
        \resizebox{\resizeSize}{!}{
            \includegraphics{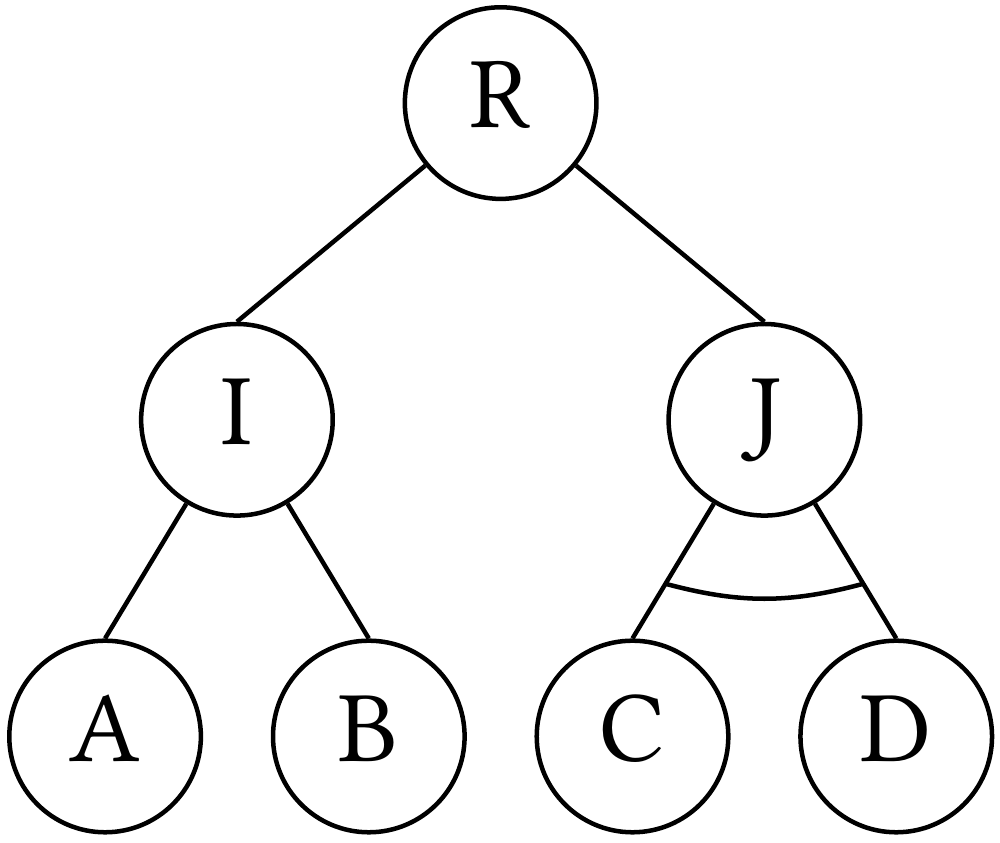}
        }
        \subcaption{Base Attack Tree}\label{sfig:base}
    \end{subfigure}
\begin{subfigure}[b]{\CEWidth}
        \centering
        \resizebox{\resizeSize}{!}{
            \includegraphics{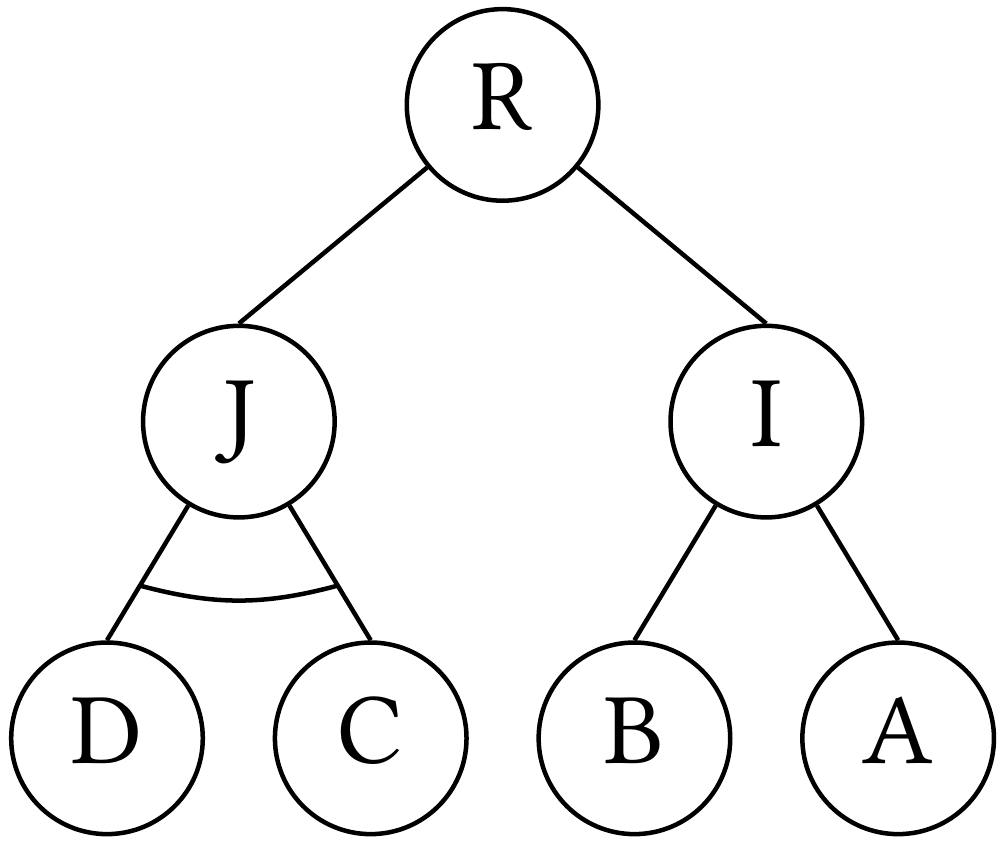}
        }
\subcaption{Order Reversed}\label{sfig:bte-order-reversed}
    \end{subfigure}
\begin{subfigure}[b]{\CEWidth}
        \centering
        \resizebox{\resizeSize}{!}{
            \includegraphics{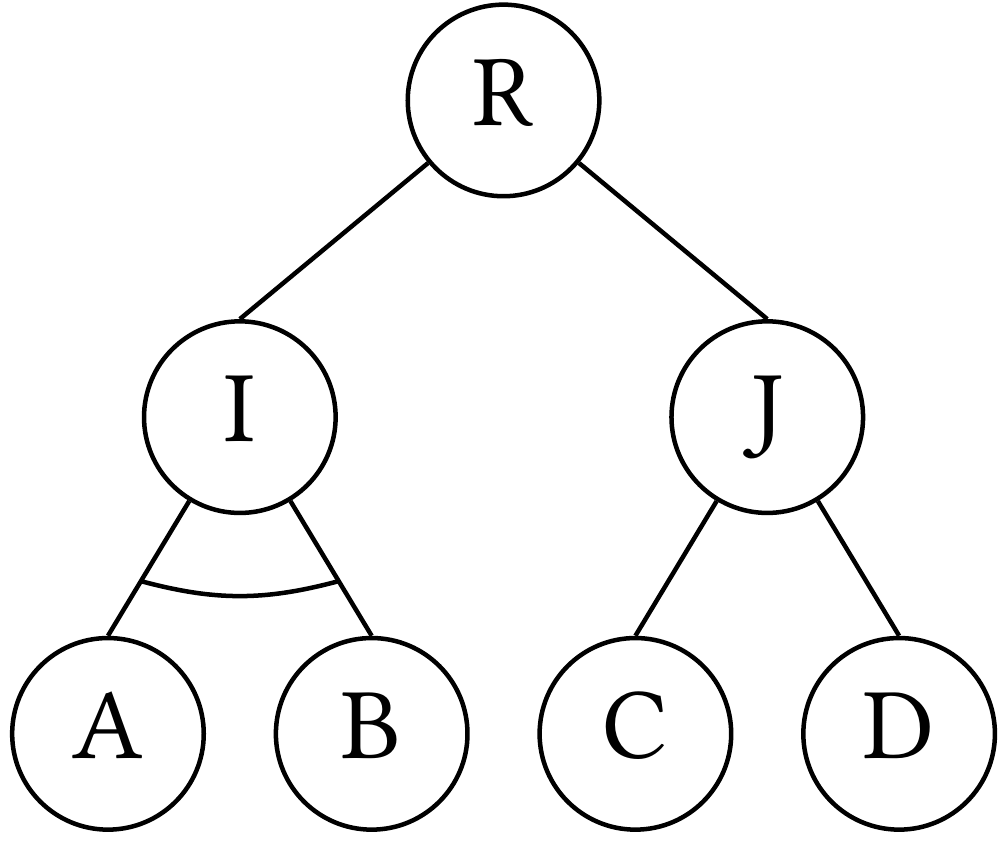}
        }
        \subcaption{Refinements Switched}
    \end{subfigure}
\begin{subfigure}[b]{\CEWidth}
        \centering
\resizebox{2.6cm}{!}{
            \includegraphics{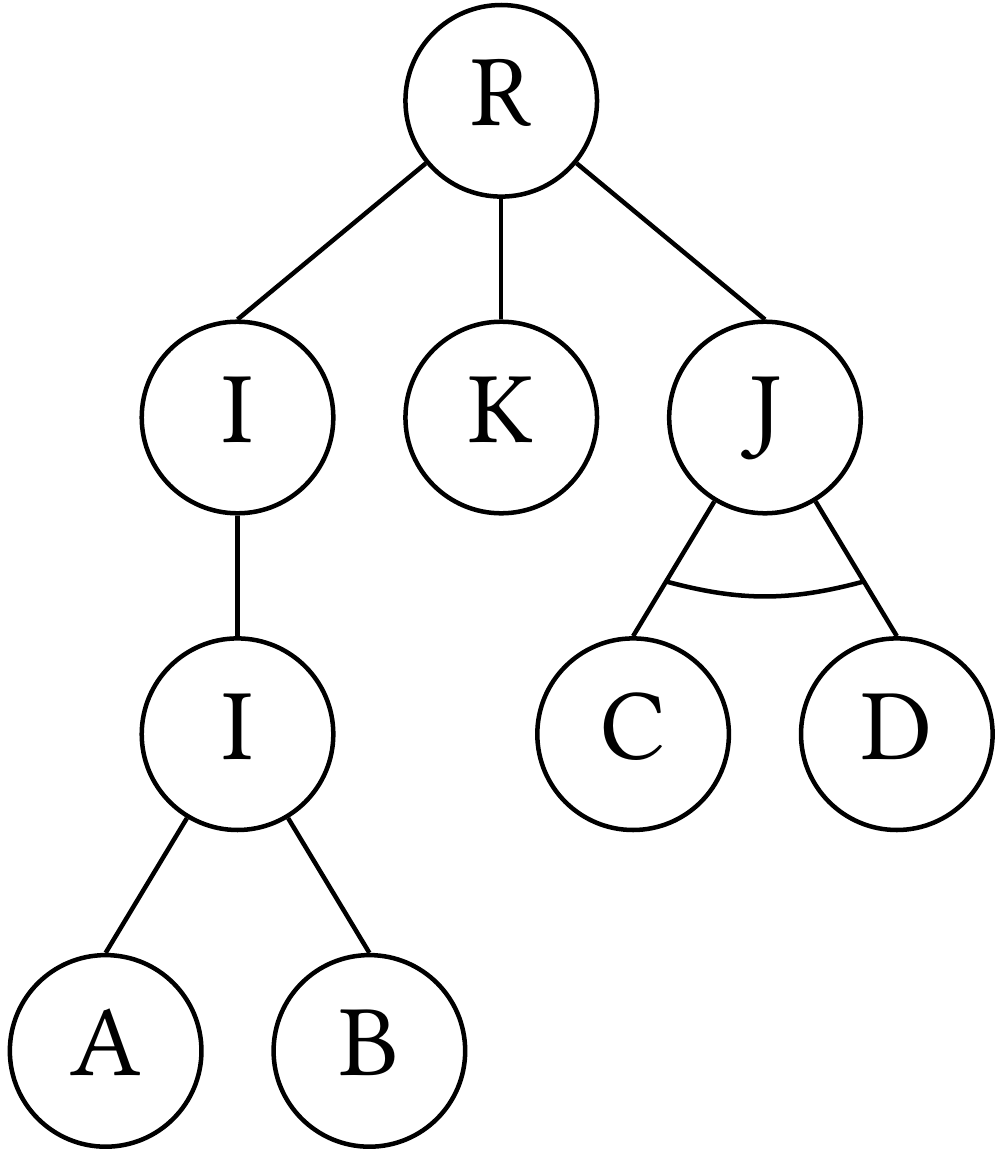}
        }
        \subcaption{\shortstack[l]{Extra\\Intermediate}}
    \end{subfigure}
\ThreeVertSpace
\begin{subfigure}[b]{\CEWidth}
        \centering
        \resizebox{\resizeSize}{!}{
            \includegraphics{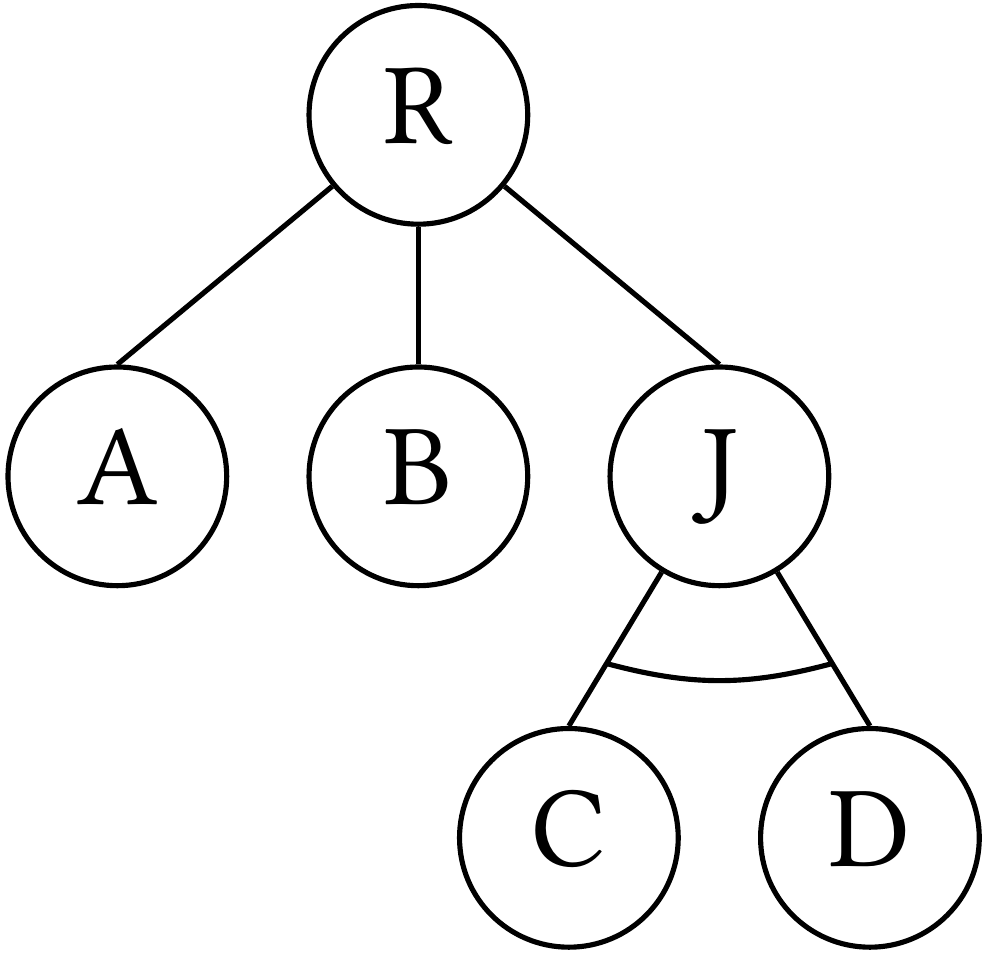}
        }
        \subcaption{Missing Intermediate}
    \end{subfigure}
\begin{subfigure}[b]{\CEWidth}
        \centering
        \resizebox{3.6cm}{!}{
            \includegraphics{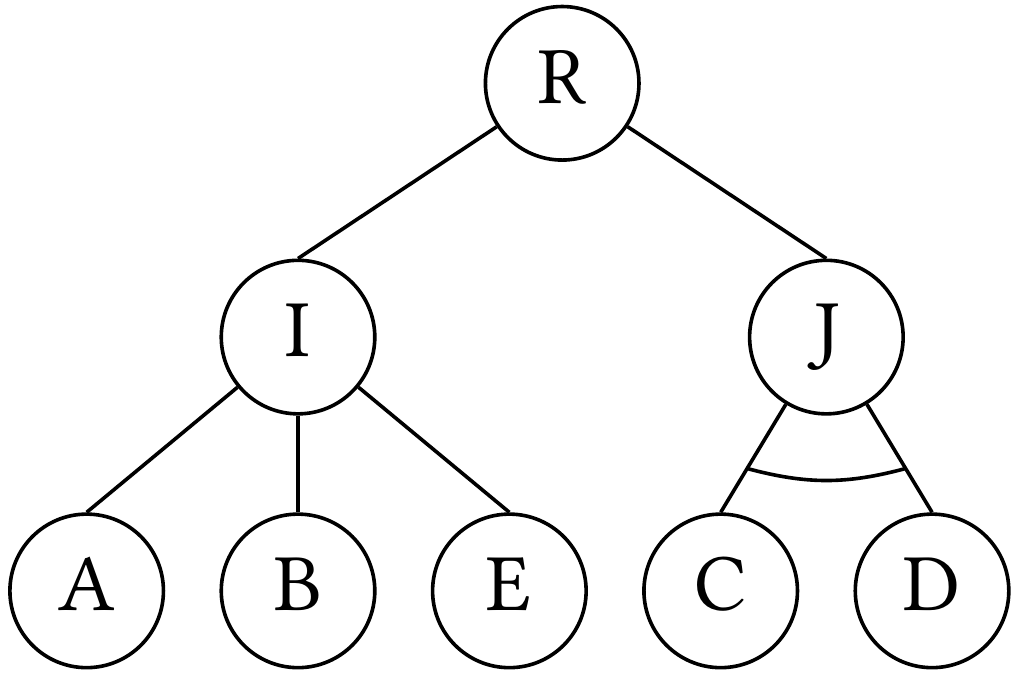}
        }
        \subcaption{Extra Leaf}
    \end{subfigure}
\begin{subfigure}[b]{\CEWidth}
        \centering
\resizebox{\resizeSize}{!}{
            \includegraphics{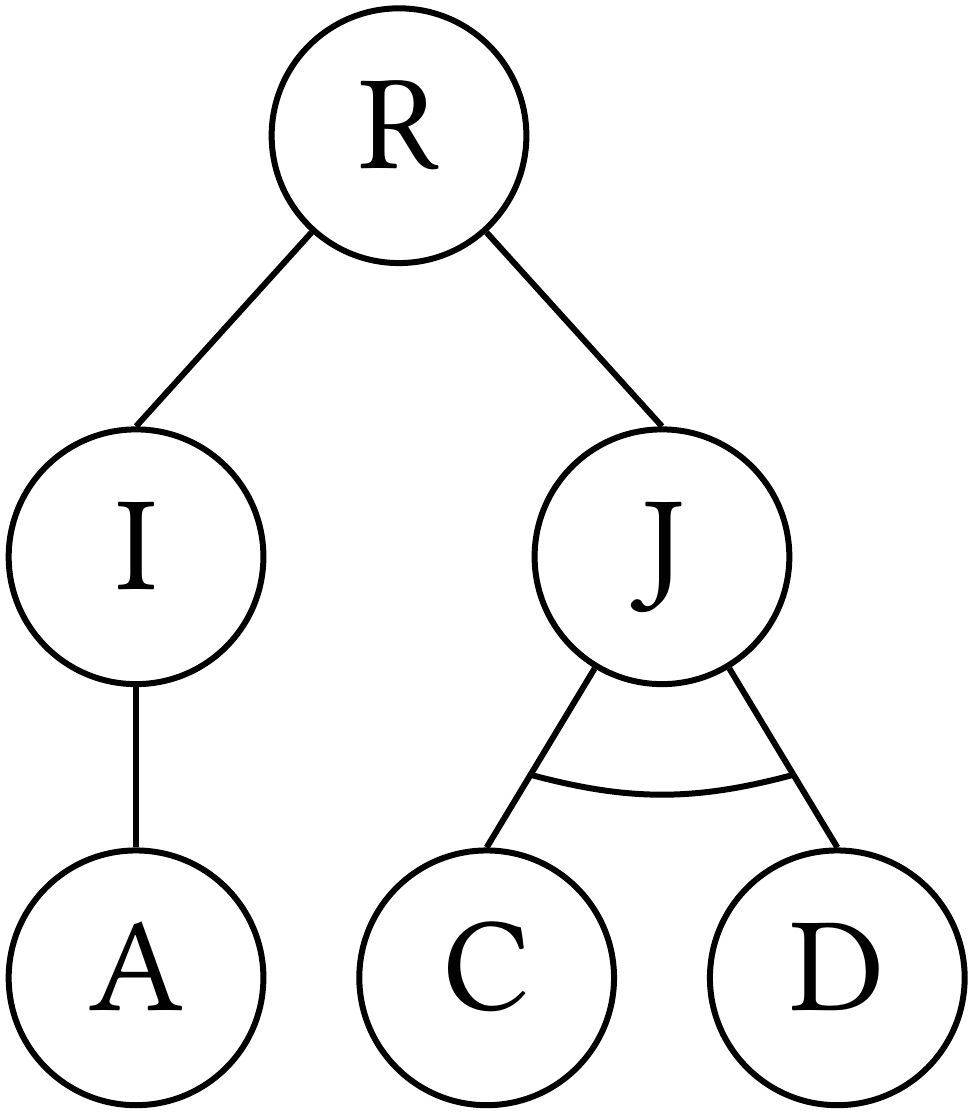}
        }
        \subcaption{MissingLeaf}
    \end{subfigure}
\ThreeVertSpace
\begin{subfigure}[b]{\CEWidth}
        \centering
        \resizebox{\resizeSize}{!}{
            \includegraphics{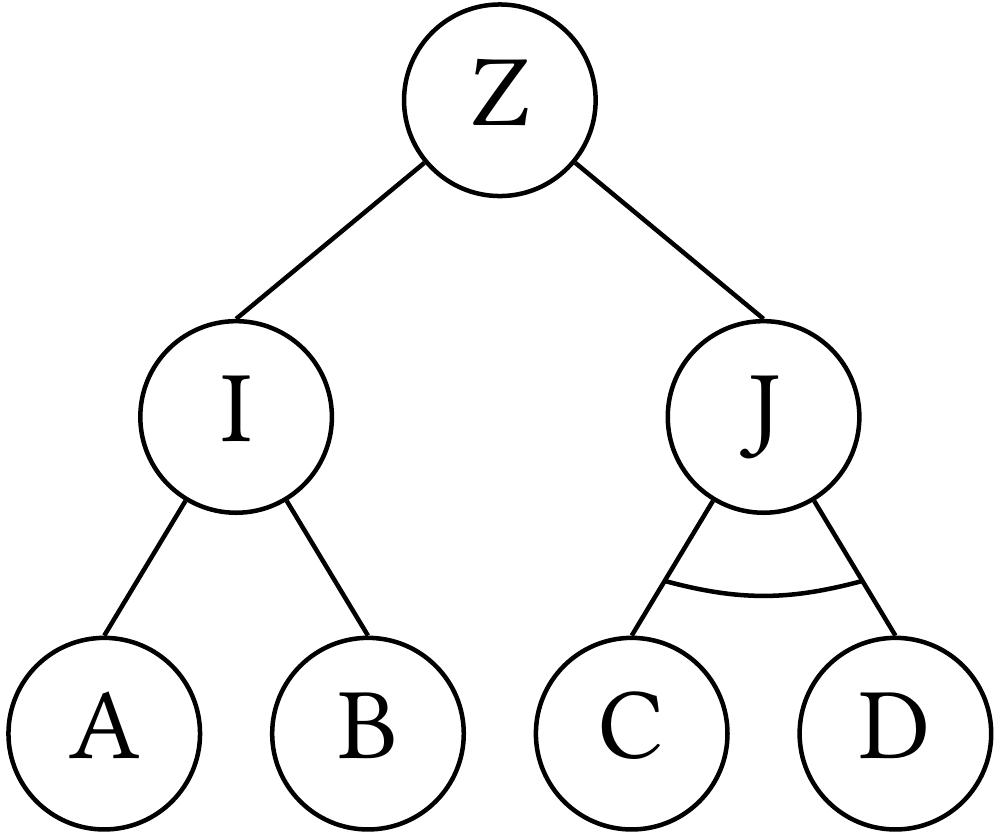}
        }
        \subcaption{Changed Root}
    \end{subfigure}
\begin{subfigure}[b]{\CEWidth}
        \centering
        \resizebox{\resizeSize}{!}{
            \includegraphics{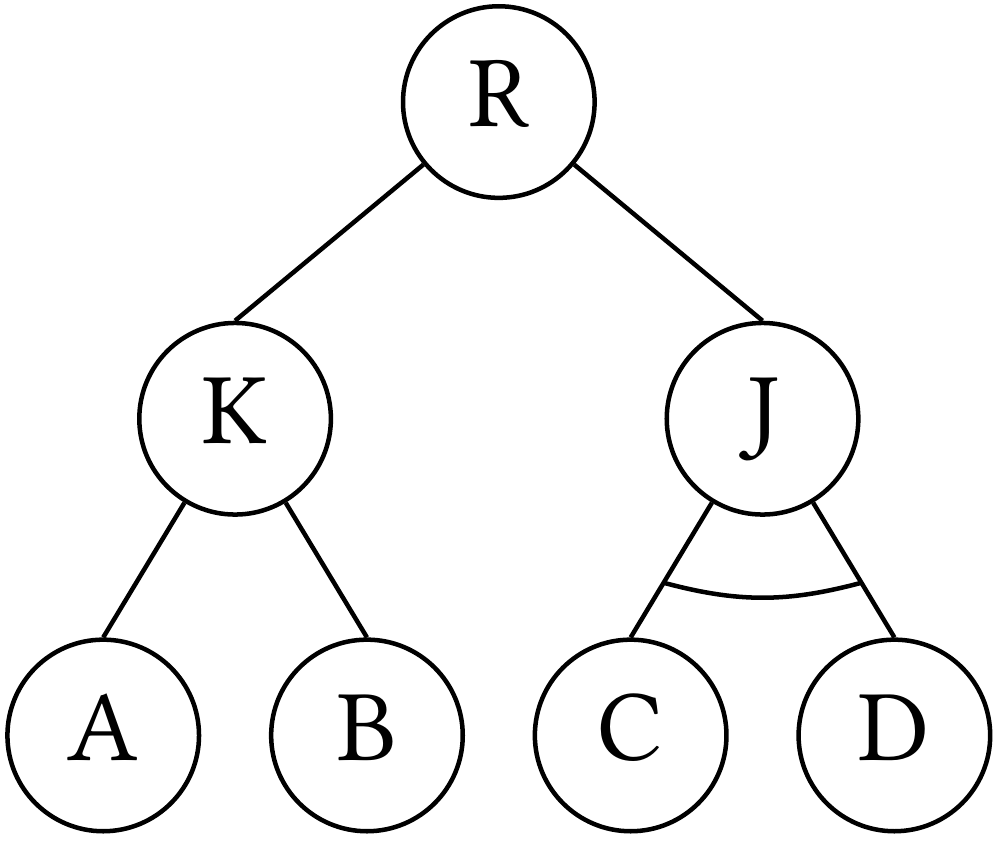}
        }
        \subcaption{Changed Intermediate}
    \end{subfigure}
\begin{subfigure}[b]{\CEWidth}
        \centering
        \resizebox{\resizeSize}{!}{
            \includegraphics{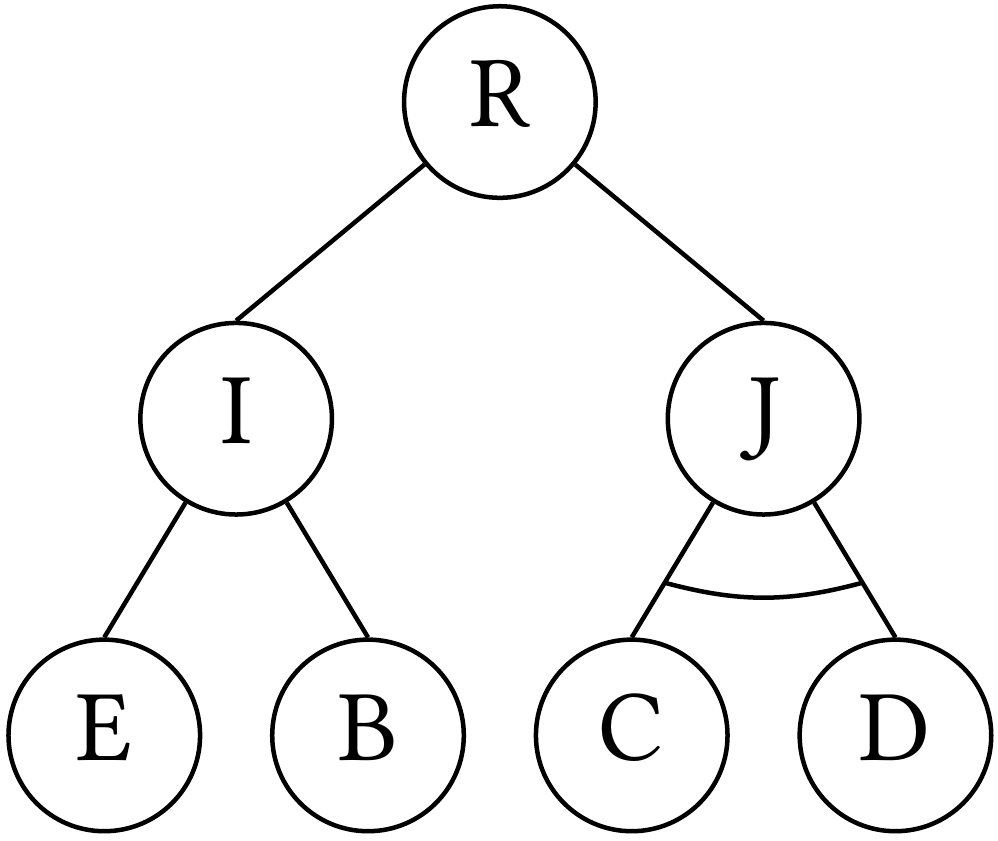}
        }
        \subcaption{Changed Leaf}
    \end{subfigure}
\ThreeVertSpace
\begin{subfigure}[b]{\CEWidth}
        \centering
        \resizebox{\resizeSize}{!}{
            \includegraphics{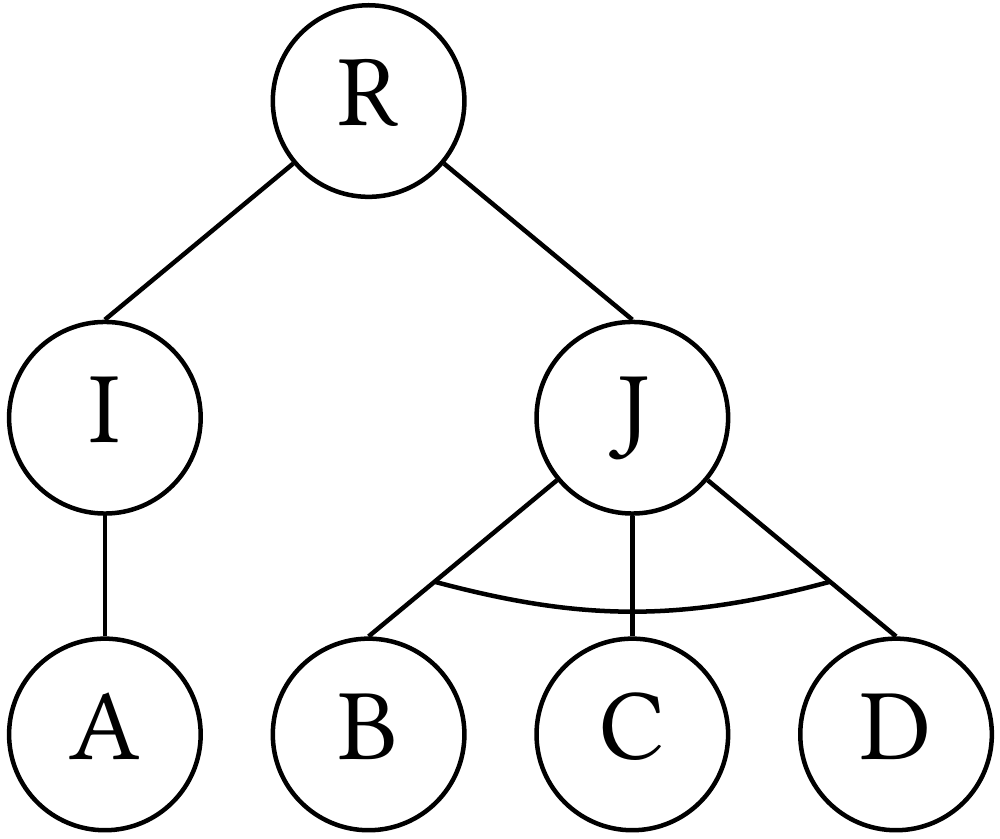}
        }
        \subcaption{Move Adjacent}
    \end{subfigure}
\begin{subfigure}[b]{\CEWidth}
        \centering
        \resizebox{\resizeSize}{!}{
            \includegraphics{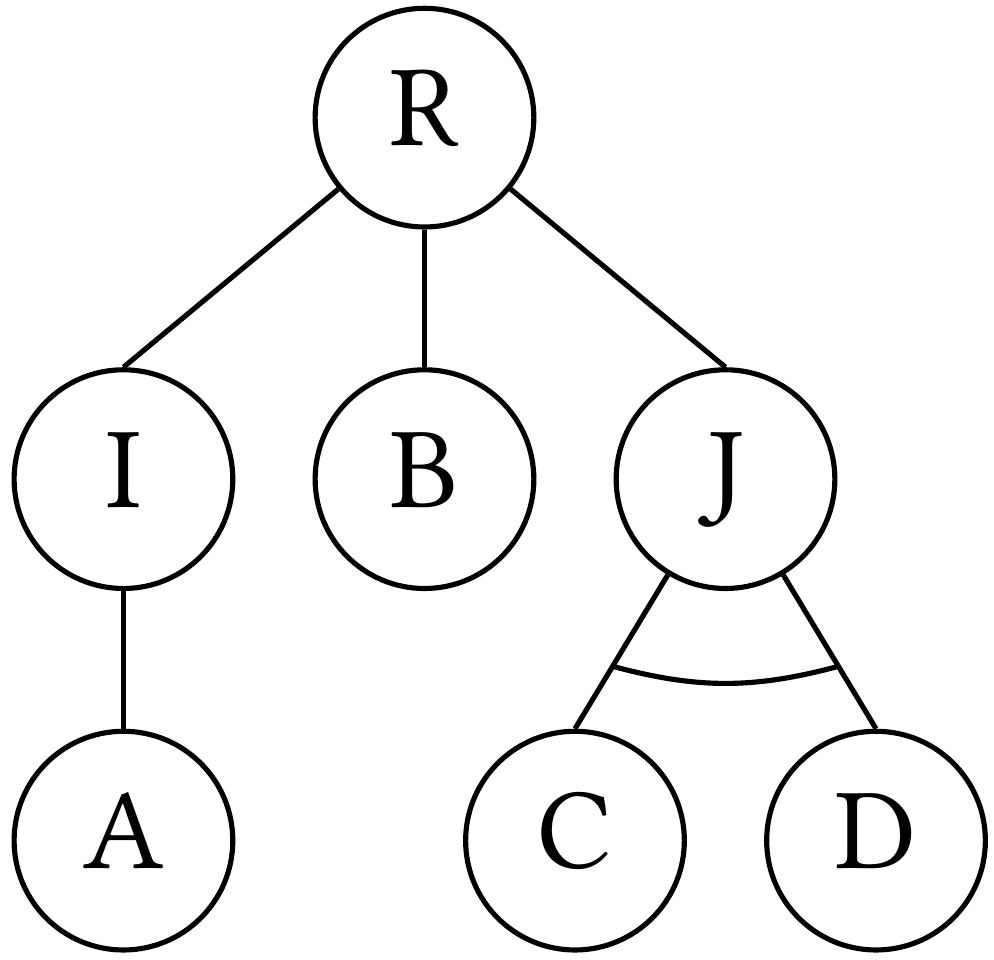}
        }
        \subcaption{Move Up}
    \end{subfigure}
\begin{subfigure}[b]{\CEWidth}
        \centering
\resizebox{2.2cm}{!}{
            \includegraphics{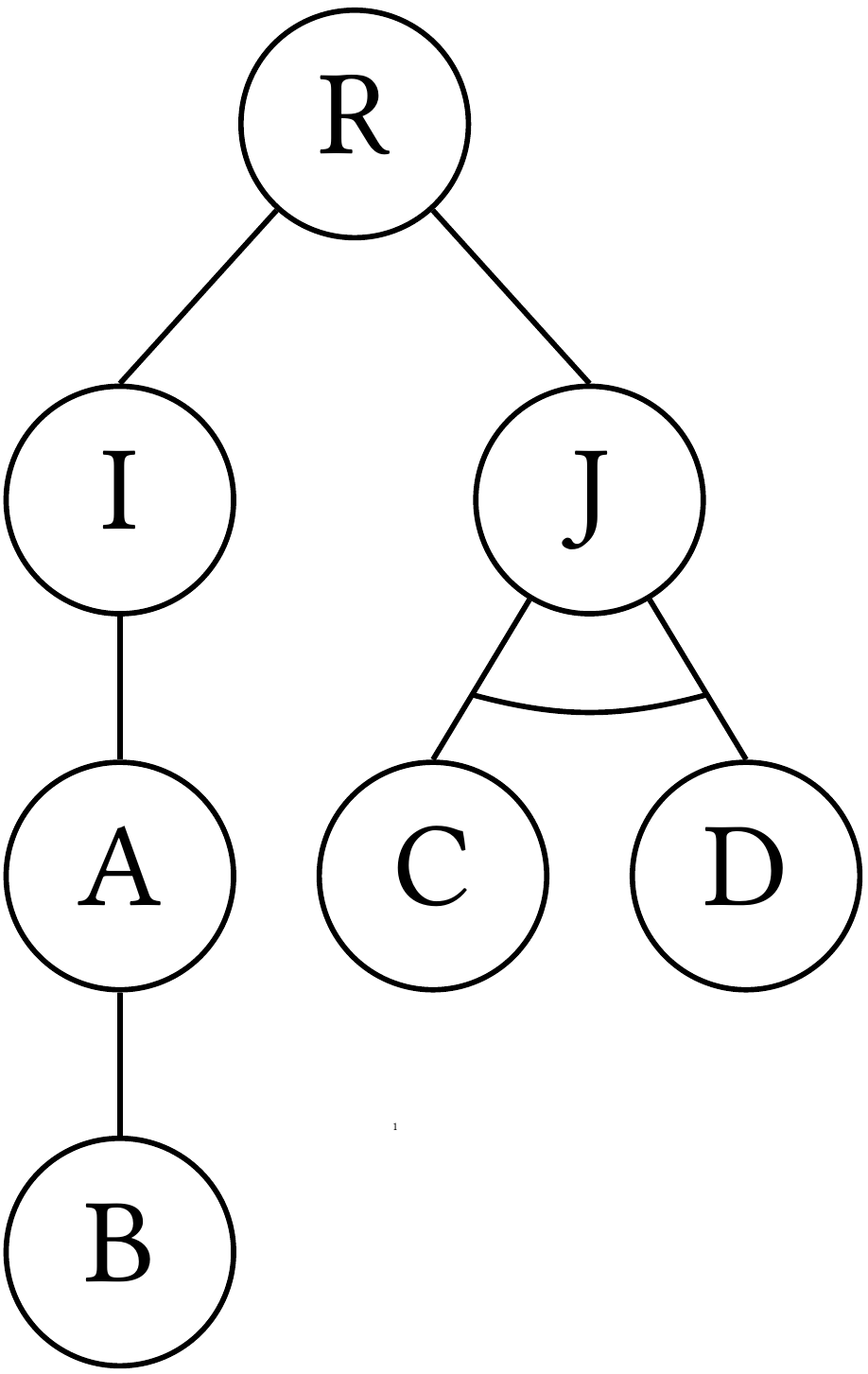}
        }
        \subcaption{Move Down}
    \end{subfigure}
\caption{Basic Transformation Examples (BTEs)}\label{fig:counterexamples}
\Description{All possible single traformations of a generic attack tree. The transformations are described by the subcaptions.}
\end{figure}
\else
\fi

\section{Results}
\label{sec:results}

We now discuss the evaluation results for the considered distances that we experimentally evaluated following the design presented in Section~\ref{sec:validation}. The code we use to perform experiments and test our results can be found at the following GitHub repository: \url{https://github.com/nschiele/ATD-TOPS-Data}.


We first evaluate the performance of the semantic similarity and the considered distance measures on the human-produced attack tree dataset (the experimental set-up and the dataset description are provided in Section~\ref{ssec:exp-validation}). Then, by examining the performance of each of our distance measures on our set of BTEs (see Section~\ref{ssec:methodology-examples}), we can evaluate the theoretical validity of each distance measure. Further, we can use the distance according to the BTEs to determine an optimal $\alpha$ for WSD.


\subsection{Evaluating alternatives to semantic similarity}
\label{ssec:method-alternatives}

As established in Section~\ref{ssec:comparing-labels}, we use SBERT with cosine similarity to establish the similarity between two node labels. This is in direct contrast to a more simplistic approach of using normalized Levenshtein distance to compare two labels. We have examined the potential effects of using normalized Levenshtein distance (with and without tokenization) to our select approach of SBERT; we present this examination here. In Table~\ref{tab:distances}, we show the results of using normalized Levenshtein distance and SBERT to compare two labels. We can see that the two methods produce very different results, with the normalized Levenshtein distance producing a value of 0 for ``door open'' and ``open door'', while the SBERT cosine similarity produces a value of 0.992. This is a clear indication that the two labels are semantically similar, while the normalized Levenshtein distance does not capture this.

\begin{figure}
    \centering
    \includegraphics[width=0.5\linewidth]{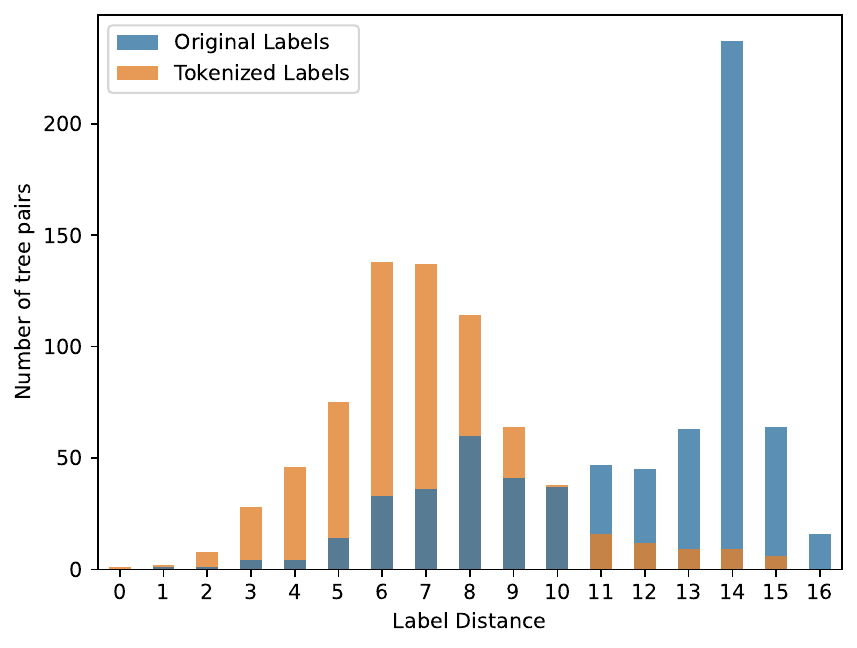}
    \caption{Pair-wise comparisons of submitted ATs for the same textual scenario ($731$ total tree pairs) using normalized Levenshtein distance to determine difference, with and without simple tokenization (lowercase, remove spaces). The collection of these tree is explained in detail in Section~\ref{ssec:exp-validation}}.
    \label{fig:tokenization_comparison}
    \Description{Exploration of how tokenization affects using string edit distance. It shows that tokenization improved performance, but not enough to be useful.}
\end{figure}

A potential mechanism to improve normalized Levenshtein distance would be to tokenize the strings. Tokenization could include shifting all letters to lowercase, removing spaces, removing prepositions, taking verb stems (removing conjugation), etc.~\cite{grefenstetteTokenization1999, a.mullenFastConsistentTokenization2018, chooStudyEvaluationTokenizer2023, mielkeWordsCharactersBrief2021}. Although this would certainly reduce the Levenshtein distance, as the selected form of tokenization would remove small differences, this does require decisions concerning which form of tokenization is appropriate and representative for the sample \cite{indurkhyaHandbookNaturalLanguage2010}. Further, it will still be the case that the tokenized Levenshtein distance does not accurately reflect the distances between two labels as a human would assess this distance~\cite{nayakDomainAdaptationChallenges2020}. 

We see in Figure~\ref{fig:tokenization_comparison} the pairwise distances of each pair of tree label sets from the AT1 set (where all tree label sets should be the same). Without tokenization, there are no two tree label sets evaluated to be identical; whereas with tokenization, four tree label sets (two distinct pairs) are evaluated to be identical. When a representative sample of label sets was independently evaluated by the authors of this work, we assessed the greatest distance between two label sets to be 4, while the greatest distance between tokenized label sets using Levenshtein distance was 15. This demonstrates that tokenization can be used to improve the accuracy of the Levenshtein distance, but it may not be the best method for comparing labels.

\if\showtables1
\begin{table}
    \resizebox{.5\linewidth}{!}{%
    \begin{tabular}{@{}llll@{}}
        \toprule
        Label 1              & Label 2             & \begin{tabular}[c]{@{}l@{}}Normalized\\Levenshtein\\ Similarity\end{tabular} & \begin{tabular}[c]{@{}l@{}}SBERT~\cite{reimers-2019-sentence-bert}\\Semantic \\ Similarity\end{tabular} \\ \midrule
        obtain personal data & obtain personnel    & 0.5882                                                                     & 0.4507                                                         \\
        obtain personal data & gather private info & 0.1176                                                                     & 0.5521                                                         \\
        break open safe      & break open door     & 0.6923                                                                     & 0.7855                                                         \\
        break open safe      & crack safe open     & 0.1538                                                                     & 0.814                                                          \\
        crack safe           & crack door          & 0.5556                                                                     & 0.6692                                                         \\
        door open            & open door           & 0.0                                                                        & 0.992                                                          \\ \bottomrule
    \end{tabular}
    }
\caption{Select examples of potential node label comparisons using both Levenshtein distance and semantic similarity. Semantic similarity is calculated using the methodology shown in Figure~\ref{fig:semanticreplacement}.}
    \label{tab:distances}
\end{table}
\else
\fi


Tokenization cannot solve the underlying issue with using normalized Levenshtein distance: comparison of the labels as strings does not consider the meaning of the labels. Strings with many similar characters may have a small Levenshtein distance while meaning completely disparate things. We demonstrate this in Table~\ref{tab:distances}. By contrast, by using SBERT to compare semantic meaning, we are able to compare the meaning of the labels, rather than the labels themselves. Thus, we used pre-trained models from Sentence-BERT~\cite{reimers-2019-sentence-bert} (built on BERT~\cite{devlin2018bert}), which create vectors to represent the semantic meaning of the input sentences.



\subsection{Determining Optimal \texorpdfstring{$\epsilon$}{ε}}
\label{sssec:finding-epsilon}

In Figure~\ref{fig:finding-epsilon}, we show our process for establishing optimal $\epsilon$ for our dataset.

\if\showimages1
\begin{figure}
    \centering
    \includegraphics[width=\linewidth]{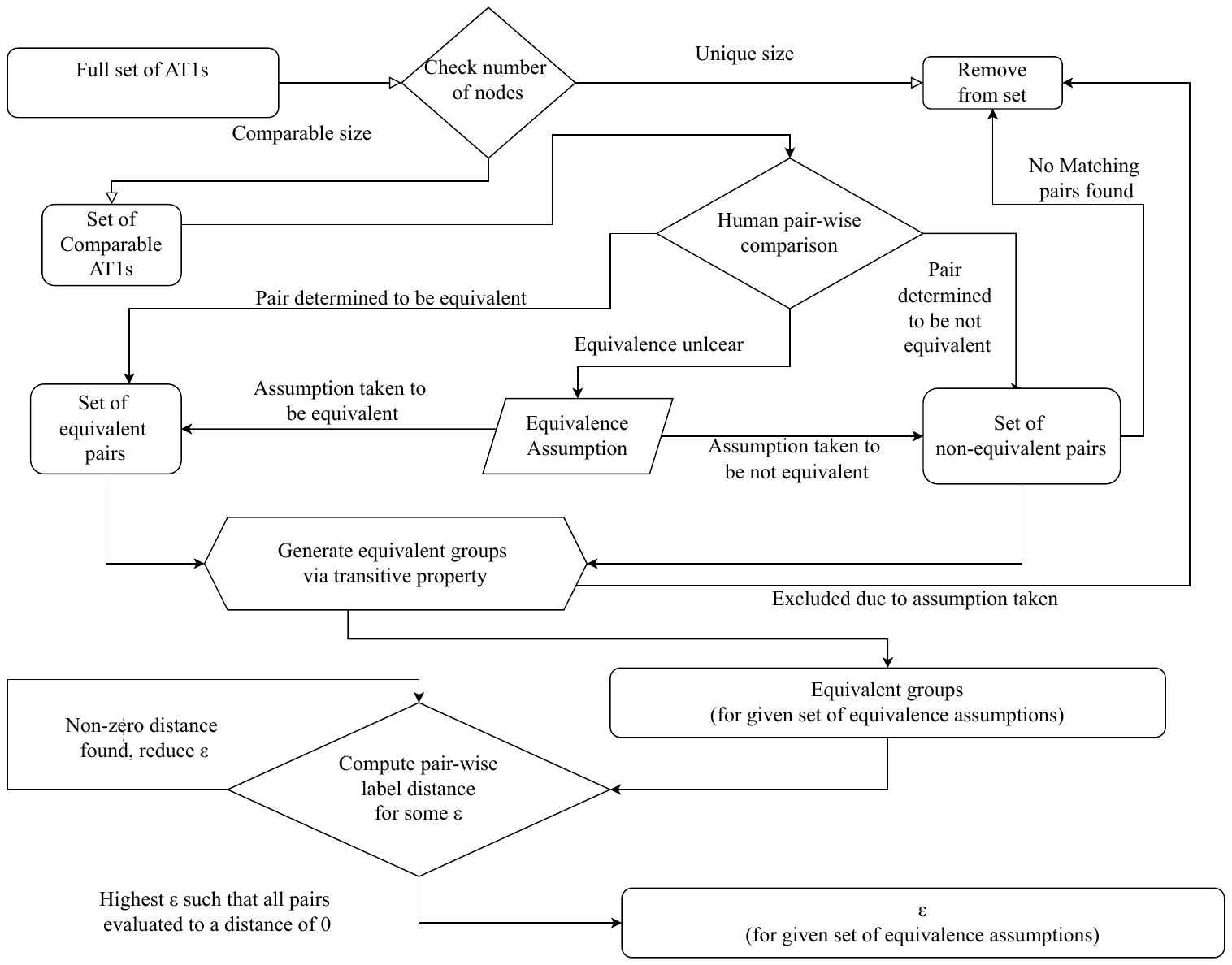}
    \caption{Flowchart showing the process described in Section~\ref{sssec:finding-epsilon}}
    \label{fig:finding-epsilon}
    \Description{Flowchart describing the steps taken to find epsilon. This is described in the text.}
\end{figure}
\else
\fi

\subsubsection{Sentence Embeddings}
\label{ssec:method-embeddings}
As we discussed in Section~\ref{sec:semantic-similarity}, SBERT is a state-of-the-art method for creating sentence embeddings~\cite{reimers-2019-sentence-bert}. We use the \texttt{SentenceTransformers} library to calculate the sentence embeddings and cosine similarity to calculate the embedding vector difference. As stated previously, our contribution is not the creation of these embeddings; additionally, our use of sentence embeddings is method agnostic, and any method for generating sentence or word embeddings could be used with our methodology. We used multiple different pre-trained SBERT models and aggregated the results to illustrate that any results we found were not the result of the use of a specific model. Models were selected to be representative of different model sizes, to ensure that the results were not found from using a specifically sized model. Additionally, we did not fine-tune any model. We used the following models: \texttt{paraphrase-multilingual-MiniLM-L12}, \texttt{all-mpnet-base}, and \texttt{all-MiniLM-L12}.

\subsubsection{Optimal Threshold}
To determine the optimal semantic similarity threshold, $\epsilon$, we follow the process shown in Figure~\ref{fig:finding-epsilon}. Based on the assignment design, the set of AT1 trees should be identical. We first manually compare a subset of label set pairs, determining if the pairs are indeed equivalent to a human. For some pairs, we made equivalence assumptions, which are documented in Table~\ref{tab:equivalence-assumptions}. This is repeated until a sufficient number of pairs of label sets is manually evaluated such that complete sets of equivalent ATs can be created via the transitive property. We then calculate the LD using a decreasing range of $\epsilon$ values until there is no distance between any of the known equivalent sets of AT1s. All other node labels either cannot be reasonably assumed to be equivalent or would in all cases be considered equivalent. Possible optimal $\epsilon$ values for a subset of possible equivalence assumptions are given in Table~\ref{tab:optimal-epsilons}. 

We have elected to accept all the ``synonymic assumptions'' (A6-8) as true, as we believe that these are simply due to word choices from our participants. We have elected to reject all the ``length assumptions'' (A1-5) as we believe that participants were including various amounts of information in their labels when labels were significantly different in length, and this indicates that the scenario was being interpreted somewhat differently. This results in an optimal $\epsilon$ of 0.7, which is the value we show in the remaining figures as a dotted line. 

\begin{table}
    \centering
\resizebox{0.95\linewidth}{!}{
\begin{tabular}{lllllcl}
\toprule
&&&\multicolumn{4}{c}{\textit{Average change in}}\\
\cmidrule[0.1pt](r){4-7}
 & \shortstack{Label 1\\$\text{ }$} & \shortstack{Label 2\\$\text{ }$} & \shortstack{Largest\\Group} & \shortstack{Average\\Group Size} & \shortstack{Total\\Comparisons} & \shortstack{Best\\$\epsilon$} \\
\midrule
A1 & submit a request for additional data other than what is needed for verification & submit a request for additional data & 5.812 & 1.220 & 54.859 & -0.029 \\
A2 & link the dataset via a request for verification & Linking the dataset & 1.938 & 1.305 & 34.234 & -0.044 \\
A3 & Unauthorized Access to Profile & Unauthorized Access & 0.000 & 0.289 & 3.234 & -0.010 \\
A4 & gaining User credentials and accessing the Profile & Gain user credentials & 0.938 & 0.235 & 10.859 & -0.006\\ 
A5 & Acquire Unauthorized Access to Profile & Acquire Unauthorized Access & 0.000 & -0.590 & 2.016 & -0.000 \\
\midrule
A6 & background data attack & get background data & 0.938 & 0.235 & 10.234 & -0.001 \\
A7 & steal personal data & obtain personal data & 0.438 & 0.495 & 8.484 & -0.068 \\
A8 & personal data extorsion & Steal personal data & 0.125 & -0.124 & 2.688 & -0.001 \\
\bottomrule
\end{tabular}
}
\caption{The eight equivalence assumptions we collected as a result of following the process shown in Figure~\ref{fig:finding-epsilon}. For each assumption, example labels are provided, along with \emph{the average change in} the number of trees in the largest equivalence group, the average equivalence group size, the total number of pair-wise comparisons, and the best $\epsilon$. These average changes are calculated by taking the difference between all cases where the assumption is taken and not taken, all else being equal. We classified assumptions A1-5 as ``length assumptions'' where, in general, we need to assume that two labels with varying amounts of information (long \& short) are equivalent. We classified assumptions A6-8 as ``synonymic assumptions'' where we need to assume that two labels with different word choices are synonymous.}
\label{tab:equivalence-assumptions}
\end{table}

\begin{table}
    \centering
\begin{tabular}{llllllllcccc}
\toprule
A1 & A2 & A3 & A4 & A5 & A6 & A7 & A8 & \shortstack{Largest\\Group} & \shortstack{Average\\Group Size} & \shortstack{Total\\Comparisons} & \shortstack{Best\\$\epsilon$} \\
\midrule
F & F & F & F & F & F & F & F & 7 & 3.5 & 30 & 0.779 \\
T & F & F & F & F & F & F & F & 11 & 4.5 & 66 & 0.652 \\
F & T & F & F & F & F & F & F & 7 & 4.75 & 54 & 0.654 \\
F & F & T & F & F & F & F & F & 7 & 3.75 & 33 & 0.645 \\
F & F & F & T & F & F & F & F & 7 & 3.75 & 31 & 0.715 \\
F & F & F & F & T & F & F & F & 7 & 3.2 & 32 & 0.756 \\
F & F & F & F & F & T & F & F & 8 & 3.75 & 36 & 0.735 \\
F & F & F & F & F & F & T & F & 7 & 3.5 & 30 & 0.779 \\
F & F & F & F & F & F & F & T & 8 & 3.75 & 37 & 0.700 \\
T & T & T & T & T & T & T & T & 18 & 6.75 & 167 & 0.573 \\
F & F & F & F & F & T & T & T & 9 & 3.6 & 46 & 0.700 \\
\bottomrule
\end{tabular}
\caption{A comparison of select different equivalence assumption selections and the effect on the size of the largest equivalence group, the average group size, the total number of pair-wise comparisons, and the best $\epsilon$ (taken to be the largest $\epsilon$ such that the pairwise distance is 0).}
\label{tab:optimal-epsilons}
\end{table}

\subsection{Effect of Semantic Similarity}

One of our contributions is the application and examination of using semantic similarity to compare node labels when comparing two trees, which would enable the use of tree distance on real-world examples, and not only on curated datasets. Of interest is how our datasets change with rising a semantic similarity limit ($\epsilon$). We expect that for each distance measure, the distance between trees will increase with a rising $\epsilon$; an increasing number of label pairs have a semantic distance above $\epsilon$, nodes that were previously marked as matching (as the cosine similarity of their semantic embeddings was below $\epsilon$) become considered as needing a ``change'' operation. In Figures~\ref{fig:semsim-at1}~and~\ref{fig:semsim-at1-2}, this is precisely the behavior we see. As described in Section~\ref{ssec:method-embeddings}, we use three pre-trained embedding models to ensure that our results are not biased to a particular model. Each solid line is the mean value across the three models of the given distance, with the deviations above and below the mean value represented by the shaded areas.

\if\showimages1
\begin{figure*}
    \centering
\begin{subfigure}[b]{.46\linewidth}
\includegraphics[width=\textwidth]{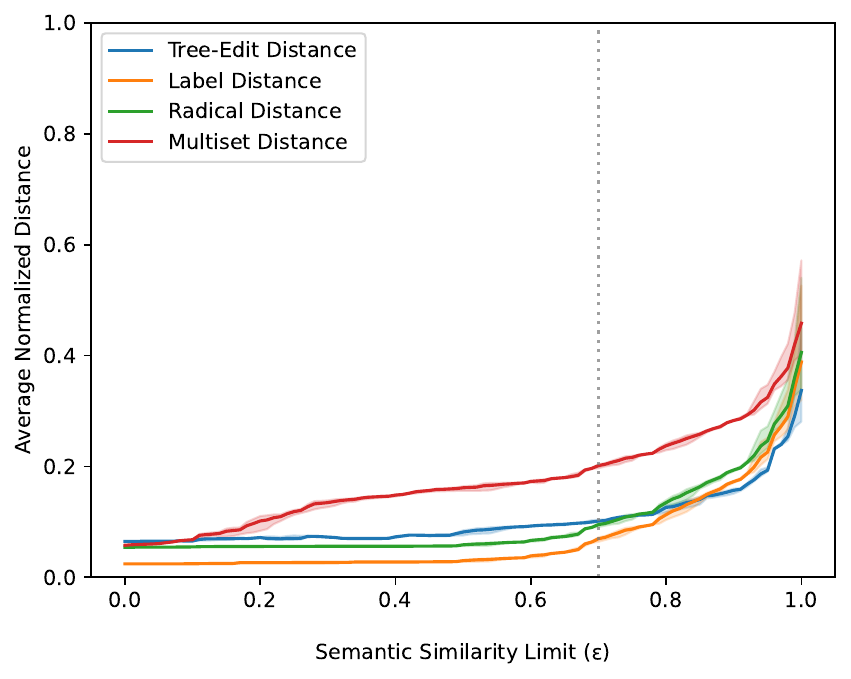}
\caption{All pairs of AT1 trees}
    \label{fig:semsim-at1}
\end{subfigure}
\begin{subfigure}[b]{.46\linewidth}
    \includegraphics[width=\textwidth]{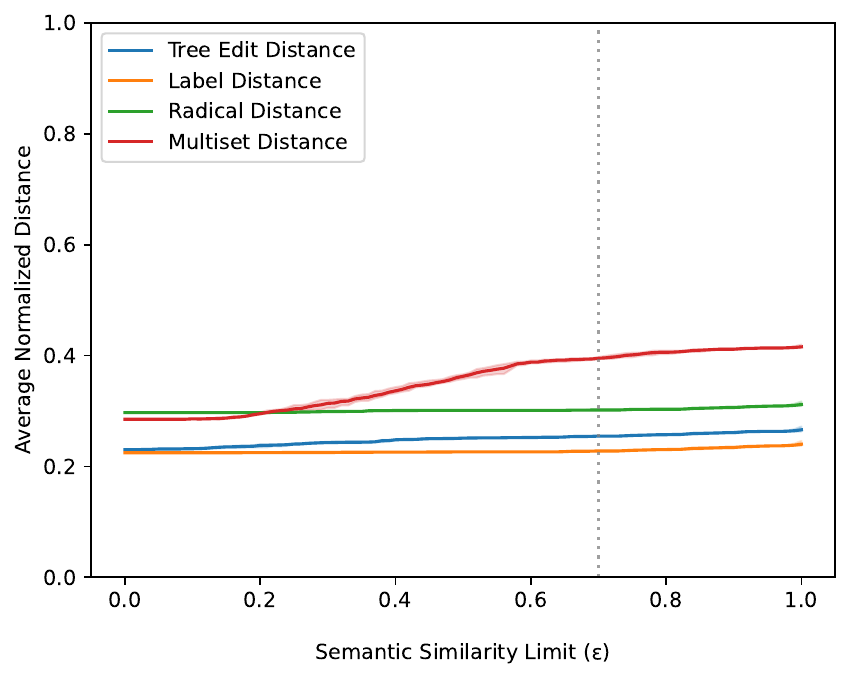}
\caption{Participant-wise comparison of AT1 to AT2}
    \label{fig:semsim-at1-2}
\end{subfigure}
\caption{Average normalized tree distance for each distance measure for semantic similarity limit ($\epsilon$) ranging from 0 to 1 with steps of 0.01.}
\label{fig:semsim}
\Description{Lineplots showing similarity measurements across different semantic similarity thresholds.}
\end{figure*}
\else
\fi

In Figure~\ref{fig:semsim-at1}, we can see the average normalized distance values for each of the distance measures plotted against an increasing semantic similarity threshold $\epsilon$. As described in Section~\ref{ssec:method-study-design}, \ATone\ was created by participants creating an attack tree from a written description (the attack tree shown in Figure~\ref{fig:tartgetAT}), which should result in identical (or nearly identical) trees. 
We expect to see distance measures that do not change with rising $\epsilon$, which for distances values below our experimentally determined optimal $\epsilon$ value of 0.7 should not change, which is largely what we see to be the case. Further, we can see that for TED and RD, the distances are largely vertically translated from LD, which is a result we would expect; the structural distance measures (TED, RD, and MSD) are built on LD with added syntactic information, and thus should have an increased normalized distance. Of note, MSD does not match the other distance measures, which would appear to indicate that MSD is measuring something other than what is expected.


In Figure~\ref{fig:semsim-at1-2}, we compare each AT1 to the AT2 generated by the same participant. We would expect all lines to be flat for all distance measures, as the only difference between AT1 and AT2 should be additional nodes, which would be unaffected by $\epsilon$. This is the exact behavior we see for LD, TED, and RD. However, for MSD, we see a different behavior of rising distance with increasing $\epsilon$. This is likely caused by multiset semantics being generated from the leaf nodes only, which means that additional nodes (and especially additional leaf nodes) fundamentally change the multiset semantics representation of a tree. This behavior suggests that MSD may be less suitable as an AT distance measure.


\subsubsection{Comparison of Operations}

One means of comparison across the different distance measures is in how they differ by operation. In TED, this is the sequence of steps taken to modify one tree into another. In the other distance measures, we can simulate these operations to create a comparable set. Unlike TED, these operations will not necessarily yield a perfectly equivalent tree after modification; however, by counting these ``operations'' per different similarity limit ($\epsilon$), we can better compare how the different distance calculations arrive at their final distance calculation. This is a means of comparison between the different distance measures that is meant to provide more insight into \emph{how} the distance measures arrive at a distance value. The difference between these as affected by the semantic similarity limit $\epsilon$ can provide insight into the behavior and validity of the distance measures.

Figure~\ref{fig:operations} compares the distribution of operations for the different distance measures. We can see that for LD, RD, and TED, the distribution of operations is relatively consistent across the different similarity limits ($\epsilon$), which is the behavior we would expect (as TED and RD are built on LD). The behavior should normalize when averaged out across the 38 different attack trees, resulting in operations plots that are similar. However, for MSD, we see a different behavior. For MSD, the distribution of operations differs fairly significantly, which is further evidence of the lack of validity of MSD as an AT distance measure.

\if\showimages1
\begin{figure*}
\centering
\resizebox{0.90\linewidth}{!}{
\begin{tabular}{lcccc}
Distance & Label Distance & Tree Edit Distance & Radical Distance & Multiset Distance \\
AT1 & \includegraphics[width=.25\linewidth]{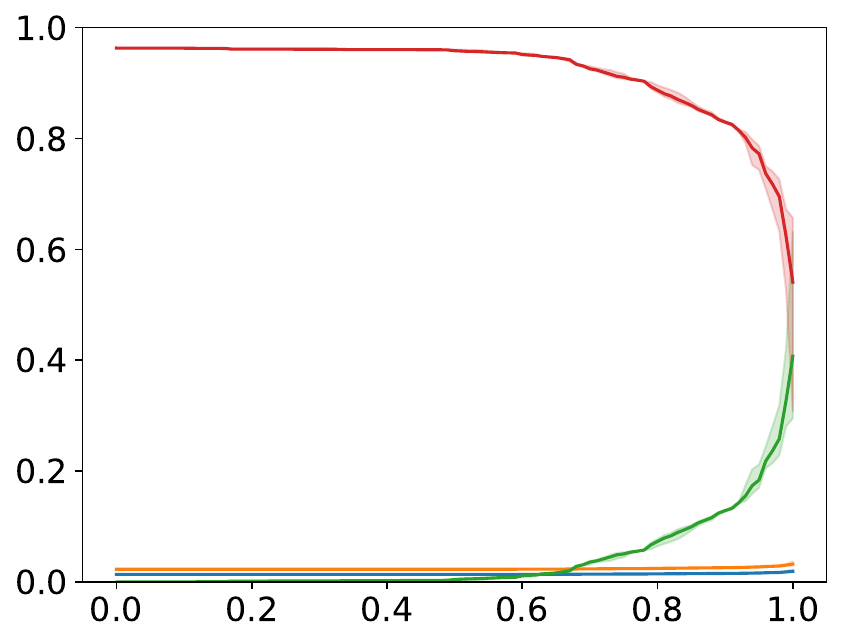} & \includegraphics[width=.25\linewidth]{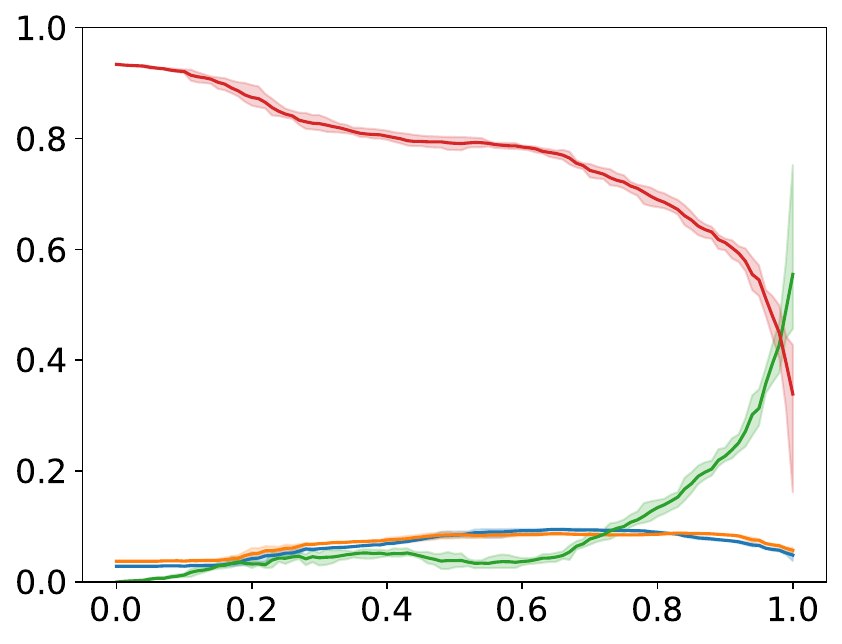} & \includegraphics[width=.25\linewidth]{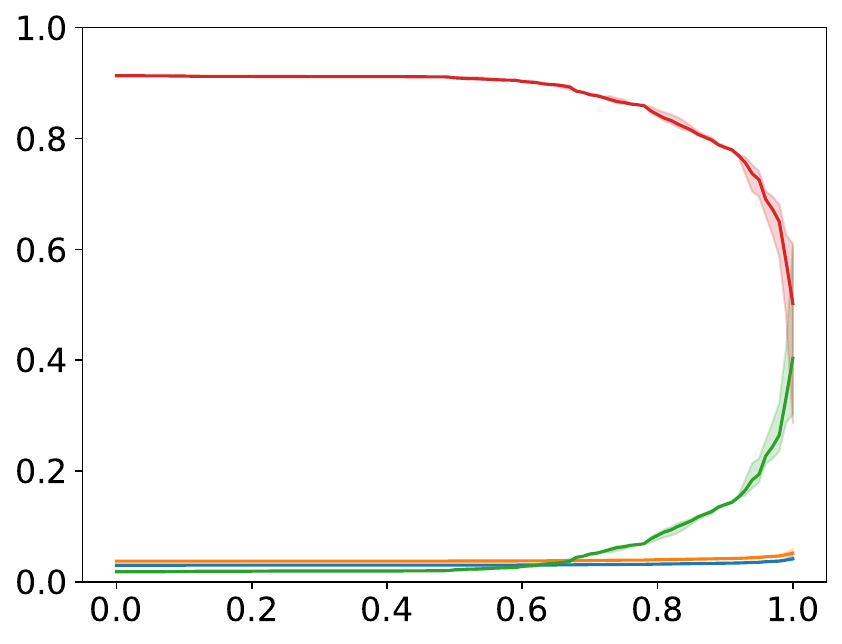} & \includegraphics[width=.25\linewidth]{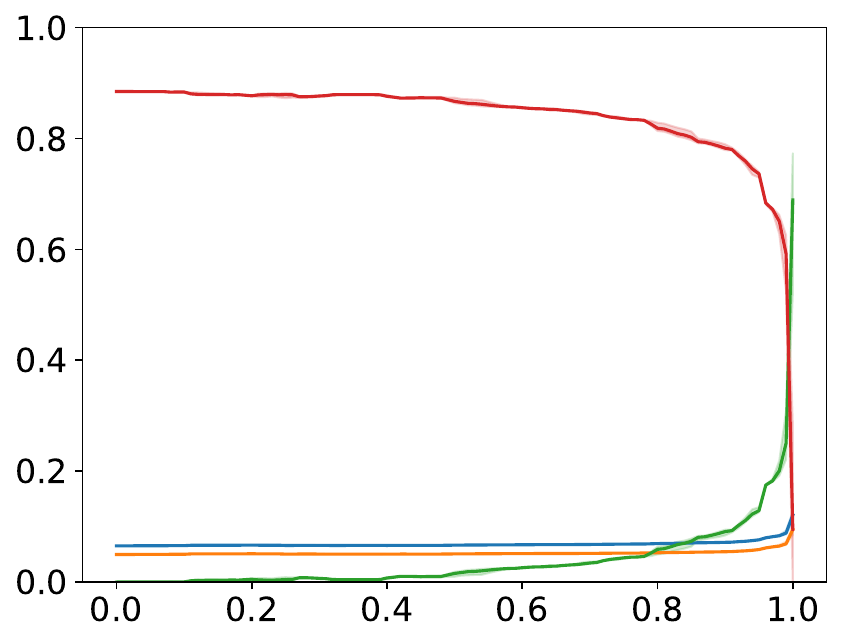} \\
AT1 vs AT2 & \includegraphics[width=.25\linewidth]{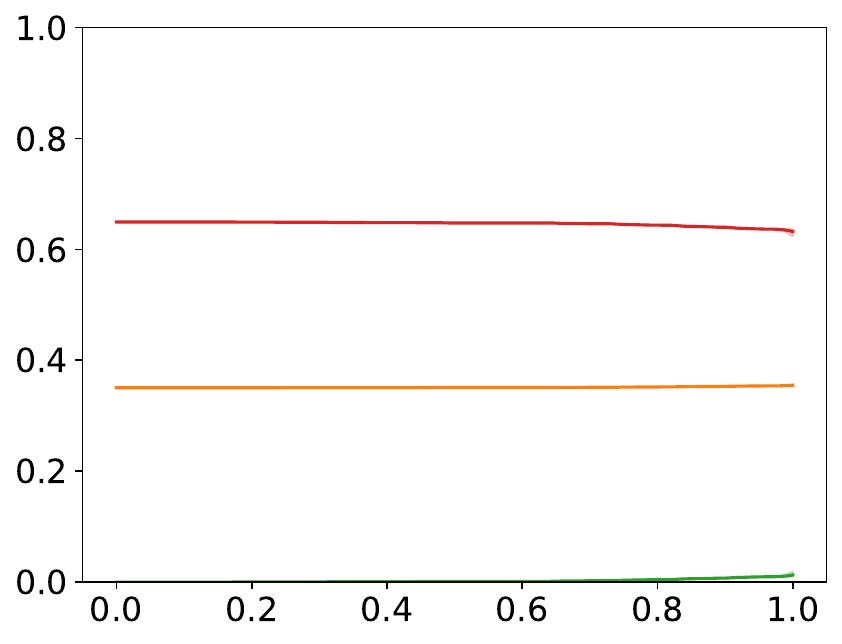} & \includegraphics[width=.25\linewidth]{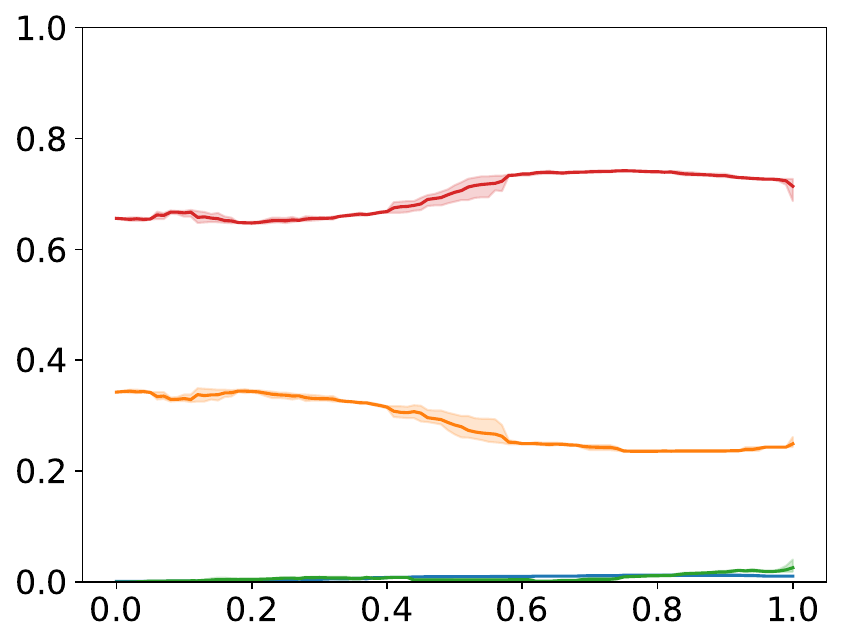} & \includegraphics[width=.25\linewidth]{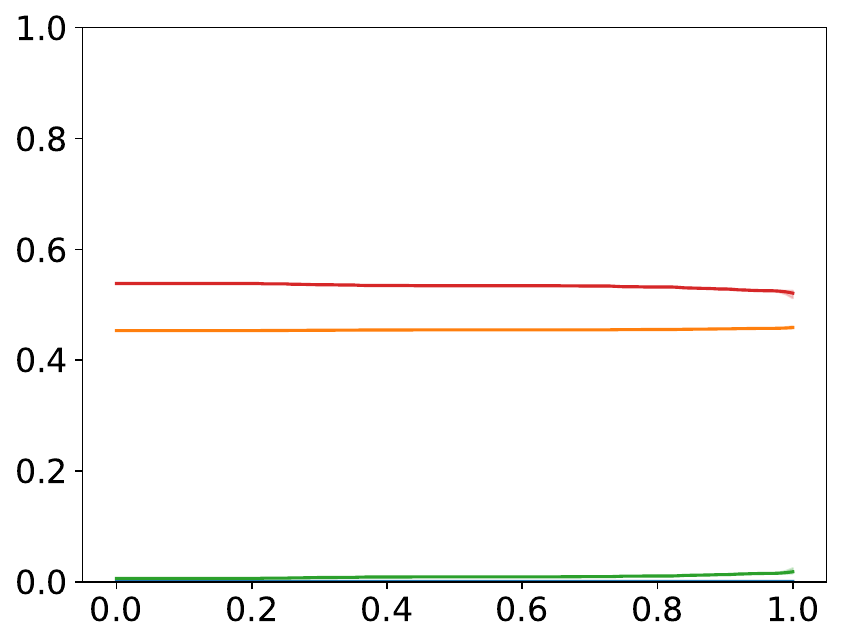} & \includegraphics[width=.25\linewidth]{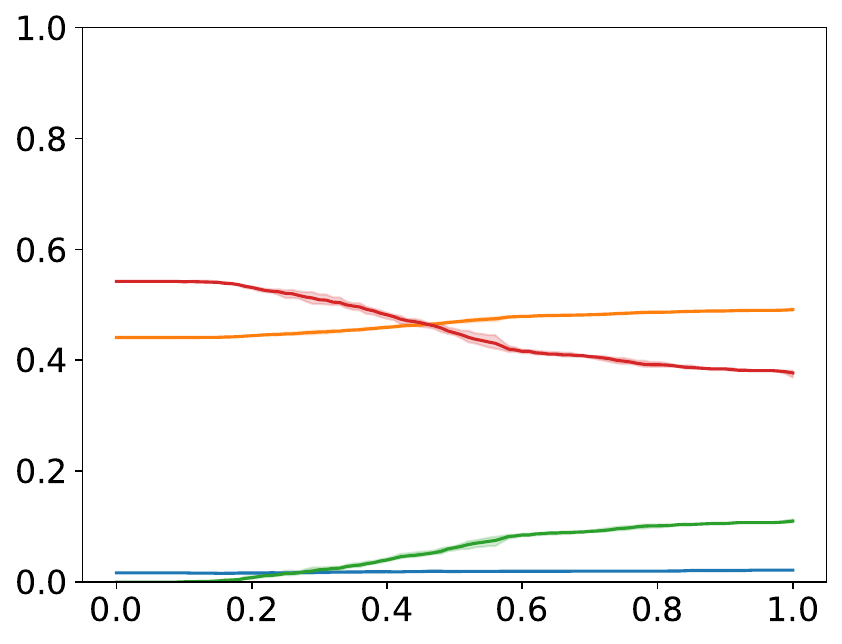} 
\end{tabular}
}
\caption{Plots showing each type of modification operation as a percentage of the total number of operations for increasing similarity limit $\epsilon$. {\color{color1} \textbf{Blue}} indicates a removal operation. {\color{color2} \textbf{Orange}} indicates an addition operation. {\color{color3} \textbf{Green}} indicates a changing operation, and {\color{color4} \textbf{Red}} indicates a matching operation. }
\label{fig:operations}
\Description{Table of lineplots shows how the different distance measures change over different semantic similarity limits.}
\end{figure*}
\else
\fi

\subsection{Theoretical Evaluation with Basic Transformation Examples (BTEs)}
\label{ssec:results-examples}

With theoretical validation, we are not examining the ability of distance measures to process ``unfiltered'' nodes (as this is addressed by the experimental validation). Here, we label all nodes with simple capital letters and set the similarity limit $\epsilon$ to 1 (effectively requiring strict equivalence to determine if nodes are equivalent).

\if\showtables1
\newcommand{\CERow}[9]{#1 & #2 & #3 & #4 & #5 & #7 & #8 & #9}
\newcommand{\WSDFootOne}{2}
\newcommand{\WSDFootTwo}{$\alpha_1$}
\newcommand{\WSDFootThree}{$\alpha_2$}
\newcommand{\WSDFootFour}{$\alpha_3$}

\begin{table*}[t]
\centering
\resizebox{0.78\linewidth}{!}{%
\begin{tabular}{lcccccccc}
\toprule
\multicolumn{1}{c}{\multirow{2}{*}{\shortstack[l]{Basic Transformation\\Example (BTE)}}} &
\multicolumn{1}{c}{\multirow[c]{2}{*}{\shortstack{Expected\\Distance}}}
&\multirow[c]{2}{*}{\shortstack{Label\\Distance} }
&\multirow[c]{2}{*}{\shortstack{Tree Edit\\Distance}}
&\multirow[c]{2}{*}{\shortstack{Radical\\Distance}}
&\multicolumn{1}{c}{\multirow[c]{2}{*}{\shortstack
        {Multiset\\Distance}}}
&\multicolumn{3}{c}{Weighted Sum Distance}
\\
&\CERow{\multicolumn{1}{c}{}}{}{}{}{\multicolumn{1}{c}{}}{\WSDFootOne}{\WSDFootTwo}{\WSDFootThree}{\WSDFootFour}
\\
\midrule
Order Reversed &\CERow{0}{0}{7}{0}{0}{1.75}{0}{3.5}{1.05} \\
Refinement Switch &\CERow{ 1}{0}{1}{1}{3}{0.5}{0.75}{0.5}{0.5} \\
Extra Intermediate &\CERow{ 1}{1}{1}{1}{0}{1}{0.85}{1}{1}  \\
Missing Intermediate &\CERow{ 1}{1}{1}{4}{0}{1.75}{1.75}{1}{2.05} \\
Extra Leaf &\CERow{ 1}{1}{1}{1}{1}{1}{1}{1}{1}  \\
Missing Leaf &\CERow{ 1}{1}{1}{1}{1}{1}{1}{1}{1}  \\
Changed Root &\CERow{ 1}{1}{1}{1}{0}{1}{.85}{1}{1}  \\
Changed Intermediate&\CERow{1}{1}{1}{1}{0}{1}{0.85}{1}{1}  \\
Changed Leaf &\CERow{ 1}{1}{1}{1}{1}{1}{1}{1}{1}  \\
Move Adjacent &\CERow{ 1}{0}{2}{2}{3}{1}{1.05}{1}{1}  \\
Move Up &\CERow{ 1}{0}{2}{2}{0}{1}{0.6}{1}{1}  \\
Move Down &\CERow{ 1}{0}{2}{3}{1}{1.25}{1.05}{1}{1.35} \\
\bottomrule
    \end{tabular}
    }
\caption{The distances provided by each of the distance measurements. The expected distance for each of these BTEs is provided in column 1\\
\footnotesize
\WSDFootTwo: We use a weighted value of $\alpha = [0.55, 0, 0.30, 0.15]$\\
\WSDFootThree: We use a weighted value of $\alpha = [0.5, 0.5, 0, 0]$\\
\WSDFootFour: We use a weighted value of $\alpha = [0.5, 0.15, 0.35, 0]$\\
\normalsize}
\label{tab:counterexamples}
\end{table*}
\else
\fi

\if\showimages1
\newcommand{\WSDWidth}{.45\linewidth}
\begin{figure*}[t]
    \centering
    \begin{subfigure}[b]{\WSDWidth}
\includegraphics[width=\linewidth]{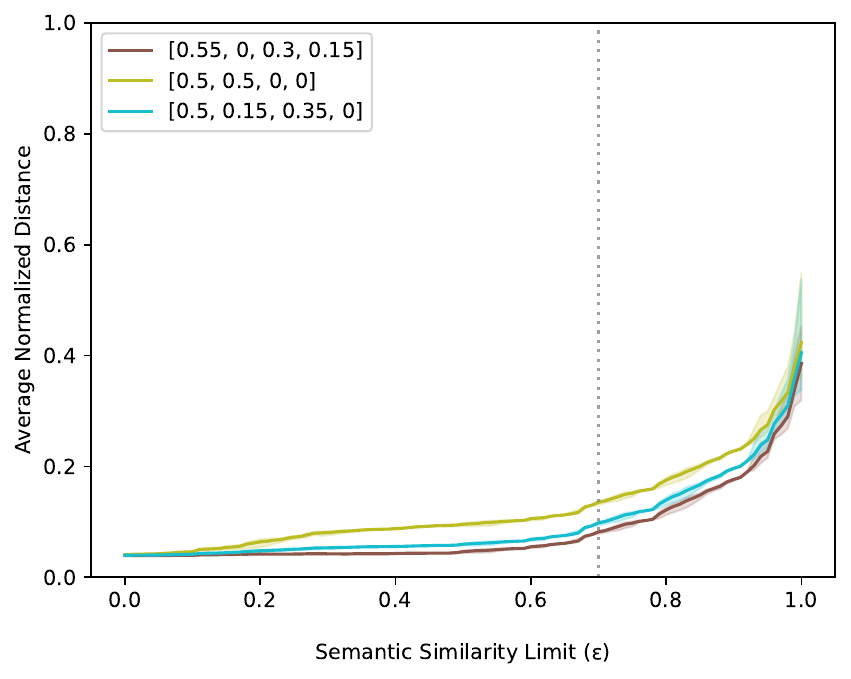}
        \caption{WSD on AT1}
        \label{fig:wsd-at1}
    \end{subfigure}
    \begin{subfigure}[b]{\WSDWidth}
\includegraphics[width=\linewidth]{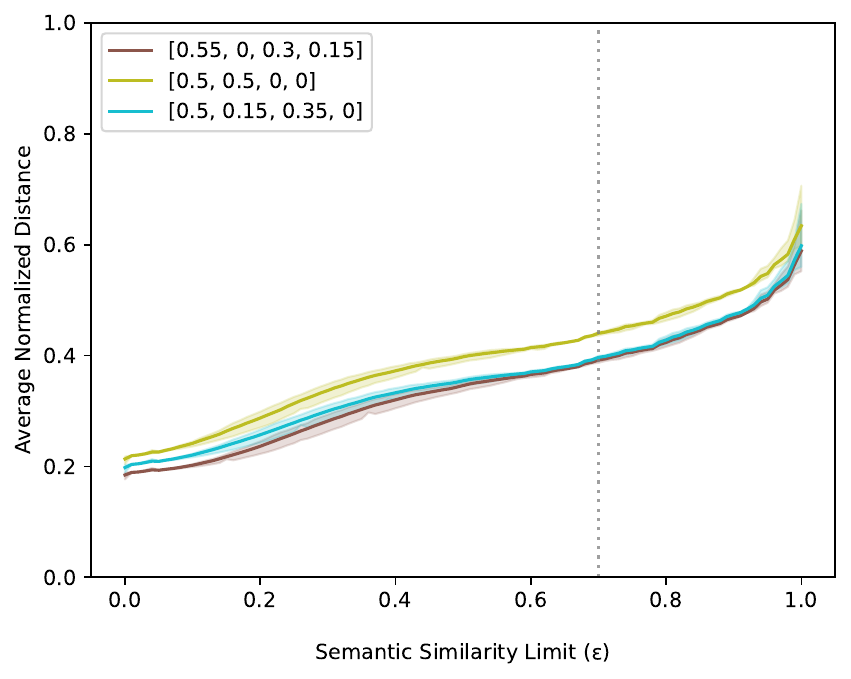}
        \caption{WSD on AT2}
        \label{fig:wsd-at2}
    \end{subfigure}
    \caption{WSD on AT1 and AT2 using the optimal $\alpha$ value established in Section~\ref{ssec:results-alpha}}
    \label{fig:wsd}
    \Description{Lineplots that are very similar to the other semantic similarity line plots.}
\end{figure*}
\else
\fi

As mentioned in Section~\ref{ssec:methodology-examples}, we have 12 examples of various basic transformation examples (BTEs) that could occur between attack trees. These are meant to assess if each of our distance measures is able to capture these changes in a manner we would consider intuitive. The absolute distance for each distance measure for each example is provided in Table~\ref{tab:counterexamples}. 
There, we describe an ``intuitive answer''. This is what we believe a practitioner would describe as the distance between the given BTEs and the base attack tree, shown in Figure~\ref{sfig:base}.

We can see that TED fails to match our intuitive answer on four BTEs, all of these with reordered nodes, as it is not capable of processing nodes in a different order. Further, TED does not match with each of the movement BTEs, as it assigns two operations (removal and addition), which results in a cost of 2. This may be correctable with a modification of the algorithm for checking if the same node is added and removed; however, this would be out of the scope of our work.

Label distance behaves exactly as expected, it perfectly matches the intuitive answer for all examples in which a label is changed, added, or removed. It does not match the intuition for any example in which the labels all remain the same, but the structure is changed in some way. LD has no mechanism to compare syntax and does not consider the structure of the tree; so long as the tree label sets do not change, neither does LD. 

Radical distance nearly matches the intuitive answer except for a few specific situations. In the example of a missing intermediate node, radical distance considers this as needing to remove three nodes (the leaf nodes in the base attack tree and the intermediate) and then re-adding the two leaf nodes. This is caused as the radicals are mapped between the two trees by their roots, which means that the radical distance has no mechanism to recognize that the two child nodes of one radical best suit another radical.
Further, RD faces the same issue as TED with moved nodes counting for a larger distance than we expect. As with TED, this is potentially correctable with an algorithm modification or a post-processing step, but this is out of the scope of our work.

Finally, MSD behaves in ways that are not at all intuitive. First, because of the construction of multisets, only the labels of leaf nodes are considered in the calculation. Some structure of the original tree exists in how the multisets are constructed, but reconstruction of the original tree is not possible \cite{mauwFoundationsAttackTrees2006}. Any modification that only applied to intermediate nodes or the root node will not be represented by multiset representations of attack trees, which likewise results in a distance of 0. For changes that affect leaf nodes, multiset semantics seem to over-represent the distance. In the 12 BTEs we created, MSD only matched the intuitive distance in 4 cases. This is further evidence that MSD is not a valid measure of distance between attack trees.

\subsection{Determining Optimal \texorpdfstring{$\alpha$}{α} for Weighted Sum Distance}
\label{ssec:results-alpha}


Looking at Table~\ref{tab:counterexamples}, we find that LD, TED, and RD all fail to match our intuitive understanding of distance on four BTEs. However, they all fail on different BTEs. MSD fails to match our intuition on seven of our BTEs.

Using our intuitive understanding of the distance values for our BTEs, we can develop three potential $\alpha$ values for WSD. The first, $\alpha_1$, is aimed at reducing the difference between our intuitive understanding of distance for each BTE and the distance calculated by WSD. We find this value to be $\alpha_1=[0.55, 0, 0.3, 0.15]$, and it results in a total deviation of 1.95 from our intuition spread across a total of eight incorrect BTEs. The second, $\alpha_2$, is aimed at reducing the total number of BTEs that are incorrect. We find this value to be $\alpha_2=[0.5,0.5,0,0]$, and it results in a total deviation of 4 from our intuition spread across a total of two incorrect BTEs. Finally, we calculate $\alpha_3$, which is intended to be an optimal value between these two extremes, optimizing for both total incorrect BTEs and total deviation from our intuition. We find this value to be $\alpha_3=[0.5, 0.15, 0.35, 0]$, and it results in a total deviation of 2.95 from our intuition spread across a total of four incorrect BTEs.

Our three optimal $\alpha$ values are shown in Table~\ref{tab:counterexamples}. Additionally, these three values are shown on AT1 and AT2 with varying $\epsilon$ in Figures~\ref{fig:wsd-at1}~and~\ref{fig:wsd-at2}. We can see that the variations between the different optimal values of $\alpha$ are relatively minute. Further, we can see the variations caused by the different models for each WSD line are smaller than for the other distances, shown in Figures~\ref{fig:semsim-at1}~and~\ref{fig:semsim-at1-2}; this suggests that WSD may be an ideal way to reduce other sources of variation. Overall, we see consistent behavior with WSD for well-chosen $\alpha$ that is in line with our expectations; most importantly, for Figure~\ref{fig:wsd-at2}, the values are consistent across the different distance measures, indicating a reliable distance measure.

\if\showimages1
\newcommand{\MDSWidth}{.45\linewidth}
\begin{figure*}[t]
    \centering
    \begin{subfigure}[b]{\MDSWidth}
\includegraphics[width=0.84\linewidth]{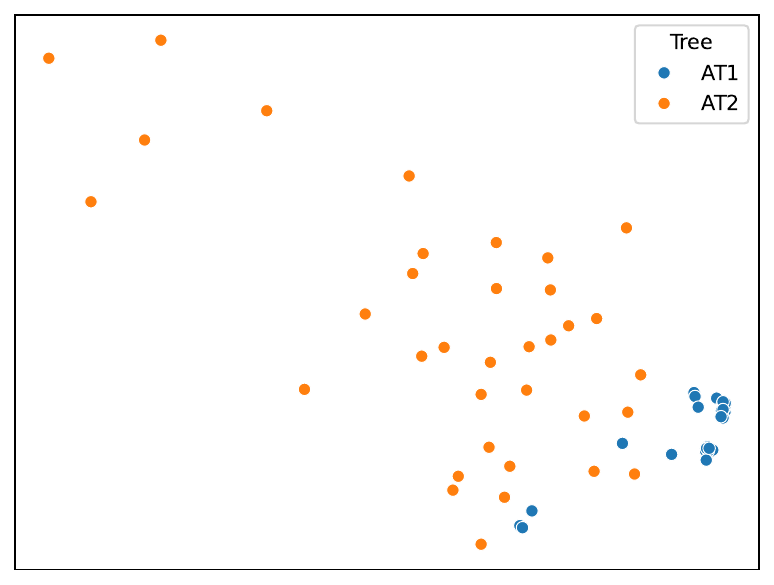}
        \caption{MDS on AT1 and AT2}
        \label{fig:mds_at1-2}
    \end{subfigure}
    \begin{subfigure}[b]{\MDSWidth}
\includegraphics[width=0.84\linewidth]{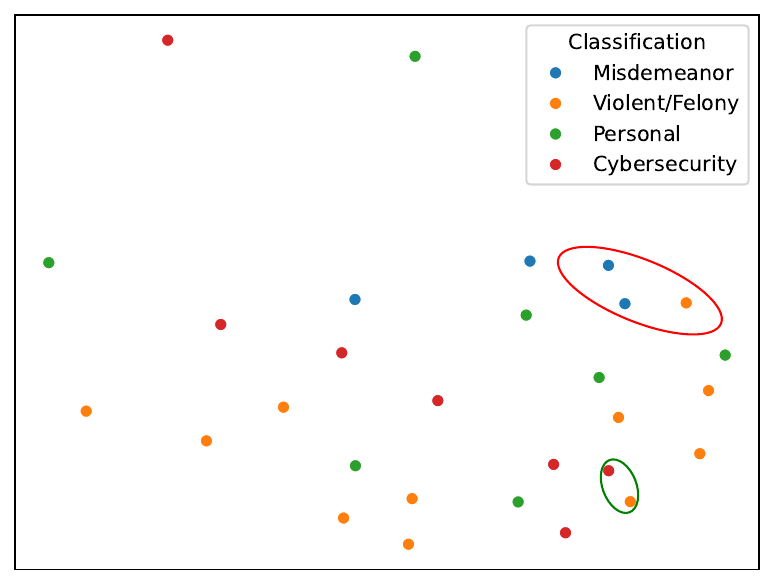}
        \caption{MDS on AT3}
        \label{fig:mds_at3}
    \end{subfigure}
    \caption{Multi-dimensional scaling applied to multiple sets of attack trees}
    \label{fig:mds}
    \Description{Scatterplots showing the distances between trees on a two dimensional plane.}
\end{figure*}
\else
\fi

\subsection{Multi-Dimensional Scaling}\label{ssec:results-mds}


Another mechanism to experimentally validate our results is by applying Multi-Dimensional Scaling (MDS). MDS is a technique used to reduce the dimensionality of a dataset~\cite{torgerson1952multidimensional}, in this case, the pair-wise distances between attack trees. We can then visually inspect the resulting 2-dimensional plot and manually compare trees that are plotted close to each other. If we find that closely plotted trees are indeed similar upon manual comparison, this is further evidence that our distance measures are valid.

Given a set of $N$ trees, a pair-wise distance between each tree can be used to generate an $N\times N$ matrix where the $i,j$ value of that matrix will represent the given distance between the $i$th and $j$th tree. We can then use multi-dimensional scaling to reduce this $N\times N$ matrix to an $N\times 2$ matrix, which can be plotted on a Cartesian plane~\cite{torgerson1952multidimensional}. This two-dimensional plot is then more understandable and can subsequently be used to aid analysis. In Figure~\ref{fig:mds_at1-2}, we can see MDS applied to both AT1 and AT2 together. We can clearly see the similarity of the AT1 trees (which should all be the same by the experiment design but are slightly different due to the noise introduced by the participants). For AT2, we can see that they differ from AT1, which is to be expected, but that they also vary from each other. If we examine close pairs of AT2s, we can see that the closer they are in the 2-dimensional graph, the more similar the trees appear to be when compared manually. 

MDS becomes more powerful when comparing completely different trees, as in the case of AT3. These trees are unique as the scenarios represented were chosen by the participants. In Figure~\ref{fig:mds_at3}, we can see the 2-dimensional representation of the distances between AT3 trees. Further, we have manually broadly classified the AT3s based on their overall topics, for comparison purposes. We highlight two examples in this plot. The green oval highlights attack trees 27 and 35, which have the root nodes of ``Home Burglary'' and ``Obtain Client Data'', respectively. As we can see, these two trees were manually classified differently; however, upon closer inspection, we find that AT3-35 has many intermediate and leaf nodes that are based around gaining physical access to computer systems, and thus is very similar to AT3-27. The red oval contains trees 5, 18, and 25 with the root nodes ``Gain Access to Festival'', ``Gain Access to Rugby Game'' and ``Steal Mona Lisa'', respectively. The similarity between AT3-5 and AT3-18 should be expected, as both concern gaining access to a physical location. AT3-25 is, judging based on the root node, very different but still contains many nodes that are similar to AT3-5 and AT3-18, as this scenario considers gaining access to a physical location. 

This is a clear example of how MDS can be used to aid in the analysis of attack trees, as an analyst would likely not be able to find this similarity without careful examination of the entire set of trees: a process that would likely take significantly longer and more prone to error than applying MDS as shown here. Additionally, as the nodes we checked in Figure~\ref{fig:mds_at3} are similar, this supports the idea that the distance measures are valid, as they are able to capture this similarity and this is translated on a 2-dimensional graph.

The distances used for MDS in Figures~\ref{fig:mds} are calculated using the $\alpha_3$ value shown in Table~\ref{tab:counterexamples}. The distances were normalized. The coordinates of the MDS graph are not provided, as the individual coordinates of a point are not meaningful; only the relative distance between points is meaningful. This is a known quality of MDS \cite{carroll1998multidimensional}. The MDS graph is generated using the scikit-learn implementation of MDS \cite{pedregosa2011scikit}.

\section{Discussion}
\label{sec:discussion}


\subsection{Label Comparison}
\label{ssec:discussion-label-comparison}

Following work in the field of Natural Language Sentence Matching (NLSM)~\cite{luMKPMMultiKeywordpair2022,wangBilateralMultiPerspectiveMatching2017}, we created a process to extract the labels from pairs of trees and compare them to calculate \emph{label distance (LD)}; a Jaccard distance-based measure between the label sets of two ATs. This distance uses semantic similarity to determine if labels are equivalent, using a well-established method of BERT to determine semantic similarity between two labels~\cite{devlin2018bert,traczBERTbasedSimilarityLearning2020}. This method is able to capture the label similarity between two attack trees. As this method does not require labels to be identical or normalized, it is directly applicable to unfiltered data. We have shown how we can compare human-created sets of labels (\SRQ{1}).

\subsection{Structural Comparison}
\label{ssec:discussion-structural-comparison}

We defined three distance measures that are built on LD and consider the refinement structure of ATs alongside the semantic similarity comparison from LD: \emph{attack tree edit distance (TED)}, \emph{radical distance (RD)}, and \emph{multiset distance (MSD)}. TED is based on the Zhang and Shasha tree edit distance algorithm~\cite{Zhang_Shasha_1989}, which is a well-established method for comparing trees. RD is a novel distance measure that considers the refinement structure of ATs, based on the refinement structure from~\cite{gadyatskayaRefinementAwareGenerationAttack2017} and the AT decomposition from~\cite{schiele2021novel}. MSD is a distance measure that considers the multiset semantics of ATs from~\cite{mauwFoundationsAttackTrees2006}. 

Finally, we proposed the \emph{weighted sum distance (WSD)}, which is a weighted sum of the previous four distance measures. This is a simple mechanism to combine the distance measures, potentially in such a way that the strengths of each distance measure are combined. With these distance measures, we have shown how we can compare the refinement structure of attack trees (\SRQ{2}).

\subsection{Validation of Distance Measures}
\label{ssec:discussion-validation}

We proposed two methods for validating our distance measures. With theoretical validation, we follow from other tree distance works~\cite{Zhang_Shasha_1989,mcvicar_sumoted_2016,pawlik_rted_2011} and propose a set of BTEs to demonstrate the behavior of our distance measures. As with previous works, our BTEs demonstrate the differing behaviors of our distance measures, showing that all distance measures behave as expected in some cases and fail to match some of our other expectations of BTEs. 
This adds to our suggestion that WSD could be a way to overcome the limitations of individual distance measures, which we attempt to do by suggesting different $\alpha$ values for WSD based on the BTEs. Overall, the BTEs are a method for showing the theoretical validity of distance measure, and we show that LD, RD, TED, and WSD are all theoretically valid in most cases, while MSD is not.

With experimental validation, we presented a repeatable experiment for creating human-generated validation data of two sets of ATs: AT1s, which should be entirely identical, and AT2s, which should be mostly identical to the AT created by the same participant. We then used this data to both find an optimal semantic similarity limit $\epsilon$ (in a thoroughly described and repeatable manner), as well as verify the distance measures we have found. We also used the freely designed trees (AT3s) to examine if the distance measures place similar trees close together, and a qualitative assessment of this mapping shows that it does, validating our distance measures. We find that LD, RD, TED, and WSD are all experimentally valid, while MSD is not. Further, by using MDS to plot WSD on a 2-dimensional plane, and qualitatively assessing trees plotted close together, we can further verified that these distance measures actually represent the difference and similarity of attack trees. We thus have shown how we can validate our distance measures for attack trees (\SRQ{3}).

\subsection{Limitations} \label{sec:limitations}

We now discuss the key limitations of this work.

\textbf{Underlying modules.}
SBERT is a popular yet not the most advanced model to establish the semantic similarity of sentences~\cite{kimSynCSESyntaxGraphbased2025,fodorCompositionalitySentenceMeaning2025,huangAddressingAsymmetryContrastive2025}. Our intention was to develop a method that can be easily iterated upon, replicated, and improved. Using SBERT as a means of establishing semantic similarity meets these goals, as it is a well-established and easily accessible tool. Still, it could be replaced with a more advanced model in the future. 
The same reasoning applies to the use of the simple Zhang and Shasha tree edit distance algorithm. While it is not the most advanced tree edit distance algorithm, it is well-established and easily accessible. In the future, it could be replaced with a more advanced or task-specific algorithm.

\textbf{Semantic similarity limitations.}
The distance measurements we evaluate with these data should be applicable to any attack trees created. However, it may be the case that the optimal $\epsilon$ and $\alpha$ may be different for attack trees from different data sets or those made by experts. We have extensively documented our methodology for finding these values, and believe that this exact process could be repeated for any new dataset.

\textbf{Limitations of our experimental validation.} As our validation method relies on human-produced attack trees, it is an open question whether these studied attack trees in sets AT1, AT2, and AT3 are representative enough of real-life, practitioner-designed attack trees. Previous work by Naiakshina~\etal\ and Karpati~\etal\ suggests that in the context of cybersecurity, students may be a sufficient proxy for practitioners~\cite{karpatiComparingAttackTrees2014, naiakshinaConductingSecurityDeveloper2020}. However, students possess limited knowledge of security concepts and thus a limited vocabulary to design sufficiently diverse and interesting attack trees. In the future, studies can try to capture ATs designed by experts following the same experimental procedure.

\textbf{Limitations of the proposed metrics.}
As our evaluation has shown, the metrics we defined are not yet a perfect solution for determining the distance between two attack trees. We propose a first step toward this goal, and we believe that our work paves the way for future research in this area. The initial tree edit distance by Tai~\cite{tai_tree--tree_1979} was not the final word on tree distance measures; instead it was the first step in a long line of research that has led to the development of many tree distance measures, with no clear definite methodology yet established~\cite{Zhang_Shasha_1989,Zhang_Statman_Shasha_1992,mcvicar_sumoted_2016,pawlik_rted_2011,paasen_-algorithm_2021}; this is described in detail in Section~\ref{sec:related-work}. We thus believe that our work will be the first step in a similar line of research for attack trees.

\section{Conclusion and Future Work}
\label{sec:conclusion}

We proposed several new approaches for defining the distance between attack trees and offered a comprehensive comparison between these measures. We have designed and demonstrated a robust and repeatable validation scheme for these measures. To the best of our knowledge, our work is the first to examine the problem of attack tree similarity. 
In the future, we plan to examine other approaches to combine the studied measures to assess the quality of automatically generated attack trees. Further, we could apply more advanced sentence matching techniques to improve these measures. \revised{Finally, these techniques could be expanded to be applied to other graphical threat models such as attack graphs or CORAS}.




\bibliographystyle{ACM-Reference-Format}
\bibliography{bibliography}

\appendix

\end{document}